\documentclass[structabstract]{aa}
\usepackage{txfonts}
\usepackage{natbib}
\bibpunct{(}{)}{;}{a}{}{,} 
\usepackage{graphicx}
\usepackage{morefloats}
\begin{document}

\title{Three dimensional interstellar extinction map towards the Galactic Bulge}


\author{B.Q. Chen \inst{1,2}
\and M. Schultheis\inst{1,3}
\and B.W. Jiang \inst{2}
\and O.A. Gonzalez\inst{4}
\and A.C. Robin\inst{1}
\and M. Rejkuba\inst{4}
\and D. Minniti\inst{5,6}
}

   \institute{Institut Utinam, CNRS UMR6213, OSU THETA, Universit\'e de Franche-Comt\'e, 41bis avenue de l'Observatoire, 25000 Besan\c{c}on, France
 e-mail: chen@obs-besancon.fr; mathias@obs-besancon.fr; annie@obs-besancon.fr
         \and
            Department of Astronomy, Beijing Normal University,
    Beijing 100875, P.R.China
    e-mail: bjiang@bnu.edu.cn
         \and
           Institut d'Astrophysique de Paris, UMR 7095 CNRS, Universit\'e Pierre et Marie Curie, 98bis boulevard Arago, 75014 Paris, France
            \and
            European Southern Observatory, Karl-Schwarzschild-Strasse 2, D-85748 Garching, Germany
  e-mail: ogonzale@eso.org; mrejkuba@eso.org
           \and
  Departamento Astronom\'iay Astrof\'isica,  Pontificia Universidad Cat\'olica de Chile, Av. Vicu\~na Mackenna 4860, Stgo., Chile
e-mail: dante@astro.puc.cl
           \and
Vatican Observatory, V00120 Vatican City State, Italy
 }

\abstract {Studies of the properties of the inner Galactic Bulge depend strongly on the assumptions about the interstellar extinction.  Most of the extinction maps available in the literature lack the information about the distance.}
 {We combine the observations with the Besan\c{c}on model of the Galaxy to investigate the variations of extinction along different lines of sight towards the inner Galactic bulge as a function of distance. In addition we study the variations in the extinction law in the Bulge. }
{We construct color-magnitude diagrams with the following sets of colors: H--Ks and J--Ks from the VVV catalogue as well as Ks--[3.6], Ks--[4.5], Ks--[5.8] and Ks--[8.0] from GLIMPSE-II catalogue matched with 2MASS. Using the newly derived temperature-color relation for M giants that match better the observed color-magnitude diagrams we then use the distance-color relations to derive the extinction as a function of distance. The observed colors are shifted to match the intrinsic colors in the Besan\c{c}on model, as a function of distance, iteratively thereby creating an extinction map with three dimensions: two spatial and one distance dimension along each line of sight towards the bulge.}
{Colour excess maps are presented at a resolution of 15\arcmin $\times$ 15\arcmin for 6 different combinations of colors, in distance bins of 1\,kpc. The high resolution and depth of the photometry allows to derive extinction maps to 10 kpc distance and up to 35 magnitudes of extinction in $\rm A_V$ (3.5\,mag in $\rm A_{Ks}$). Integrated maps show the same dust features and consistent values with the other 2D maps. Starting from the color excess between the observations and the model we investigate the extinction law in near-infrared and its variation along different lines of sight.}
 {}


\maketitle

\titlerunning{3D extinction maps for Bulge}
\authorrunning{Chen et al.}

\section{Introduction}

Interstellar extinction is a serious obstacle for the interpretation of  stellar populations in the Galactic bulge. It shows a non homogeneous clumpy  distribution  and is extremely high  towards the Galactic Center.

Several studies of cumulative extinction at the distance of the inner Galactic Bulge have been made:  \citet{schultheis1999}  used the DENIS near-infrared data set in combination with theoretical RGB/AGB isochrones from \citet{bertelli1994}.  They showed that for an appropriate sampling area the observed sequence matched well with the isochrone with appropriate reddening.  The maximum extinction that could be reliably derived from the $\rm J$ and $\rm Ks$ data available from DENIS was about $\rm A_V=25^{m}$.   They found that in some areas, presumably with the highest extinctions, there were no J band counterparts for sources detected in $\rm Ks$.   For these regions, only lower limits to the extinctions could be obtained. \citet{dutra2003} obtained similar results using the same technique with 2MASS data. \citet{gonzalez2011} mapped the extinction along the minor axis of the Bulge based on the Vista Variables in the Via Lactea (VVV) ESO public survey data \citep{saito2012} using  the red clump giants. \citet{gonzalez2012} presents the full extinction map over the entire VVV data set covering 315 sq. degrees using the red clump technique. This map extends beyond the inner Milky Way bulge, to also higher Galactic latitudes, and is the most complete study extending up to $\rm A_{V} \sim 35^{m}$, much higher extinction values than most previous studies. Recently, \citet{nidever2012} determined high-resolution $\rm A_{Ks}$ maps using GLIMPSE-I, GLIMPSE-II and GLIMPSE-3d data
 based on the  Rayleigh-Jeans Color Excess ("RJCE") method. 

However, most of the studies  are restricted to 2D extinction  maps. \citet{drimmel2003}  built a theoretical large scale three dimensional Galactic dust extinction map.  \citet{marshall2006} presented  a 3D model  of the extinction properties of the Galaxy by using the 2MASS data and the stellar population synthesis model of the Galaxy, the so-called  Besan\c{c}on model of the Galaxy \citep{robin2003}. However, their study is limited by the confusion limit of 2MASS in the galactic Bulge region and the limiting sensitivity of 2MASS in high extincted regions ($\rm A_{V} > 30^{m}$). Recently, \citet{robin2012} did a major improvement in the Besan\c{c}on model by adding a  bar component which  results in an excellent agreement by comparing 2MASS color-magnitude diagrams as well as in the metallicity and radial velocity distribution (see \citealt{babusiaux2010}, \citealt{gonzalez2012}, \citealt{uttenthaler2012}  ). We will use this updated model to build up a new 3D extinction model using the GLIMPSE-II  data which maps the Galactic Plane within $\pm 10$ degrees in longitude from the Galactic Center with a wavelength coverage between 3.6 to 8.0\,$\mu$m allowing to map the highest extincted regions as shown in \citet{schultheis2009}. In addition, we  use the recently published VVV data with a pixel size nearly 10 times smaller and a sensitivity in Ks of 3--4 mag deeper than 2MASS  to map the 3D extinction in the near IR.

The infrared color-color diagram J--Ks vs. K--[mid-IR] has been used  by \citet{jiang2006}  and \citet{indebetouw2005} to determine the extinction coefficients $\rm A_{mid-ir}$ along different lines of sight. \citet{gao2009}  traced the extinction coefficients based on the data from GLIMPSE survey and found systematic variations  of the extinction coefficients with Galactic longitude which appears to correlate with the location of the spiral arms. \citet{zasowski2009} found  a Galactic radial dependence of the extinction law in the mid-IR using GLIMPSE data while \citet{nishiyama2009} determined the extinction law close to the Galactic Center between 1.2 to 8.0 $\mu$m.

In this paper, we first introduce our data set in Sect.~2. We discuss in Sect.~3  the  Besan\c{c}on stellar population synthesis  model together with the recent improvements. In Sect.~4 we describe the method to determine the extinction coefficients and in Sect.~5 the method to derive the 3D-extinction. We finish with Sect.~6 with the determination of the extinction coefficient part as well as the conclusion (Sect. 7).

\section{The data set}

\subsection{The GLIMPSE-II survey}
The Galactic Legacy Infrared Mid-Plane Survey Experiment (GLIMPSE-II) surveyed about 20 square degrees of the central region of the Galactic Inner Bulge using the Spitzer Space Telescope  \citep{werner2004} equipped with  the Infrared Array Camera (IRAC) \citep{fazio2004}. It surveyed approximately 220 square degrees of the Galactic plane at four IRAC bands [3.6], [4.5],
 [5.8], and [8.0], centered at approximately 3.6, 4.5, 5.8 and 8.0 $\mu$m respectively. The GLIMPSE-II data, covers a range of longitudes from -10$^{\circ}$ to 10$^{\circ}$ and a range of latitudes  from $\pm$1\,deg to $\pm$2\,deg depending on the longitude. The GLIMPSE-II coverage does not include the Galactic center region $\mid l \mid$ $\le$ 1$^{\circ}$  and $\mid b \mid$ $\le$ 0.75$^{\circ}$, which was observed by the GALCEN program \citep{ramirez2008}.

We use here the  point source archive (GLMIIA, \citealt{churchwell2009}). The GLMIIA catalogs consist of point sources with a signal to noise higher  than 5 in at least one band and less stringent selection criteria than the Catalog (see \citealt{churchwell2009} for more information). It contains the entire GLIMPSE II survey region including the GALCEN data and GLIMPSE I data at the boundary of the surveys (at l=10$^{\circ}$ and l=-10$^{\circ}$). Moreover, these catalogs  were also merged with   2MASS  resulting in a seven bands catalog: three 2MASS bands and four IRAC bands. The photometric uncertainty of the  GLIMPSE-II data is typically better than 0.2 \,mag and the astrometric accuracy  better than 0.3\arcsec.

\subsection{The VVV survey data}

\begin{figure}[!htbp]
   \includegraphics[width=9.3cm]{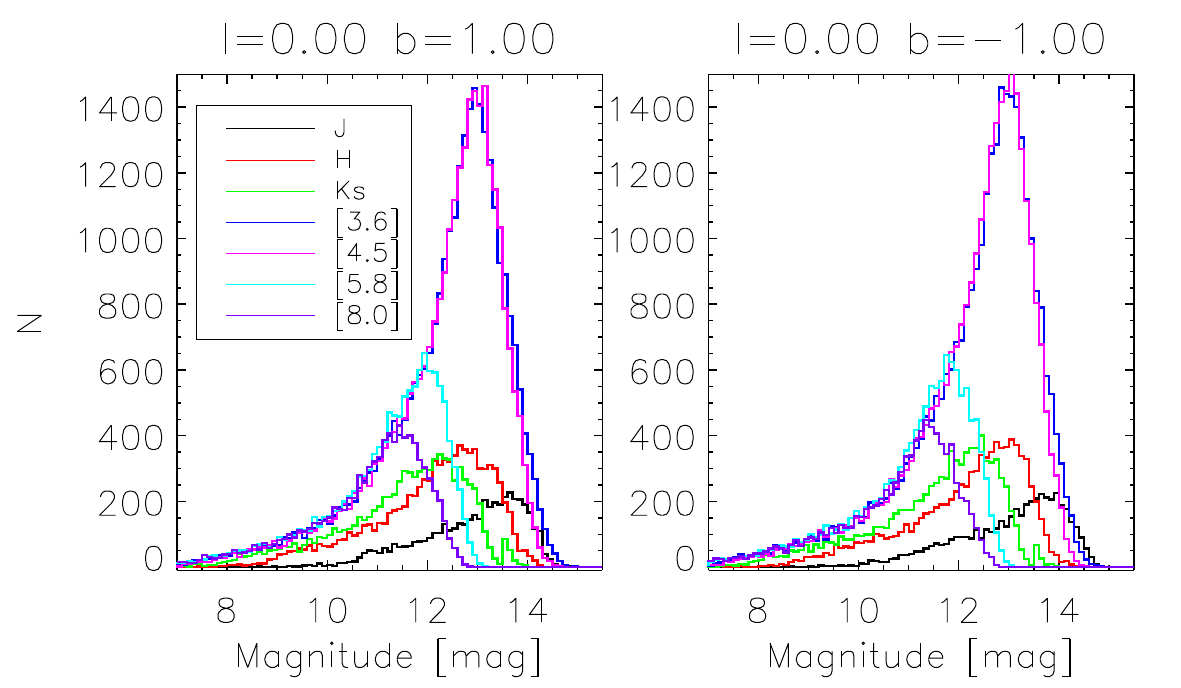}
\caption{Star counts of GLIMPSE II data of two example fields for the 2MASS bands and IRAC bands. Each field is 15' $\times$ 15'.}
\label{fig1}
\end{figure}
\begin{figure}[!htbp]
   \includegraphics[width=9.3cm]{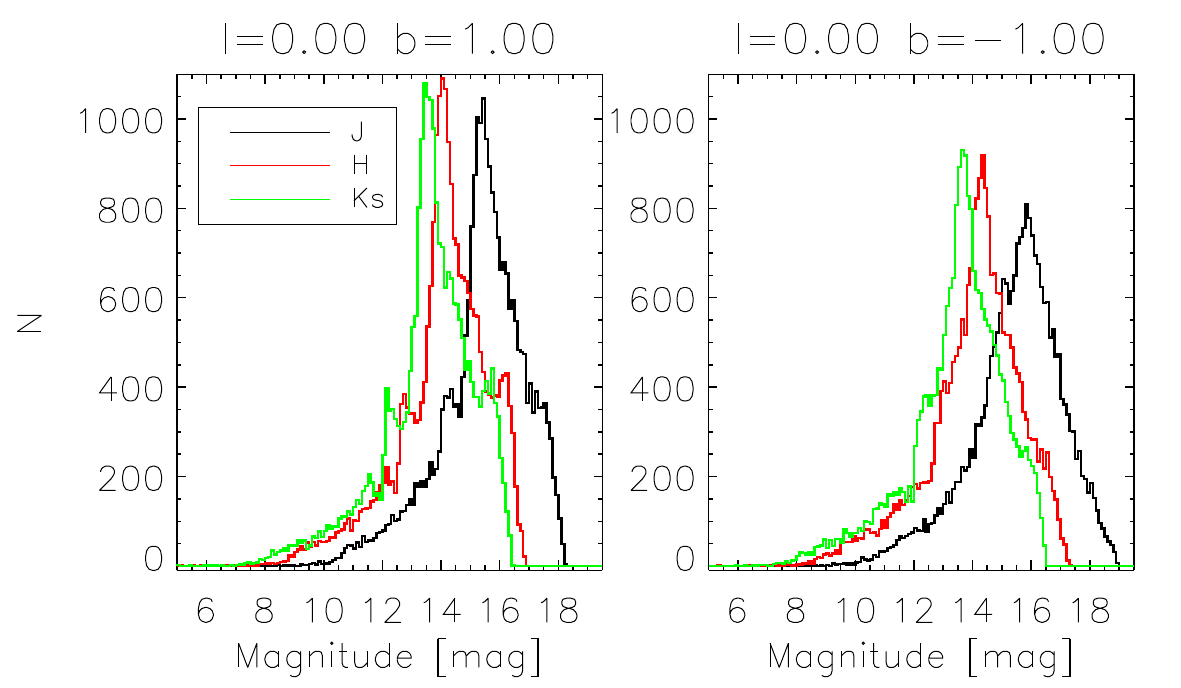}
\caption{Star counts of VVV survey data of two example fields for the 2MASS bands. Each field is 15' $\times$ 15'.}
\label{fig2}
\end{figure}

The other data set comes from the Vista Variables in the Via Lactea (VVV) ESO VISTA near-IR public survey. Observations were carried out using the VIRCAM camera on VISTA 4.1m telescope (Emerson \& Sutherland 2010) located at ESO Cerro Paranal Observatory in Chile. The five year variability campaign of the survey, which started observations during 2010, will repeatedly observe in Ks band an area of $\sim 564$ deg$^2$, including a region of the Galactic plane from 295$^{\circ}$ $<$ l $<$ 350$^{\circ}$ to -2$^{\circ}$ $<$ b $<$ 2$^{\circ}$, and the Bulge spanning from -10$^{\circ}$ $<$ l $<$ 10$^{\circ}$ and -10$^{\circ}$ $<$ b $<$ +5$^{\circ}$. During the first year of observations, the complete survey area was fully imaged in five bands: Z, Y, H, J, and Ks. In this work, we make use of the J, H and Ks catalogs for the same field of the GLIMPSE-II data from the 314 sq.deg coverage of the Galactic bulge. For a complete description of the survey and the observation strategy we refer to \citet{Minniti2010}. A detailed description of the data products used in this work, corresponding to the first VVV data release, can be found in \citet{saito2012}. 

A description of the single-band photometric catalogs, produced by the Cambridge Astronomical Survey Unit (CASU), and the procedure to build the multi-band catalogs can be found in \citet{gonzalez2011}. These catalogs were used in \citet{gonzalez2012} to construct the complete 2D extinction map of the Bulge and are the same ones used for the present study. We provide here only a brief summary of the procedure. J, H, and Ks catalogs produced by the CASU are first matched using the SLITS \citep{taylor2006} code based on the proximity of sky positions. Since VVV photometry is saturated for stars brighter than Ks=12 \,mag, 2MASS photometry is used to complete the bright end of the catalogs. Therefore, VVV catalogs are calibrated into the 2MASS photometric system. As described in Gonzalez et al. (2011), this calibration zero point is obtained by matching VVV sources with those of 2MASS in the magnitude range of $\rm 12^{m} < Ks < 13^{m}$. This magnitude range was selected to avoid the saturation limit of VVV and the faint 2MASS magnitudes with poor photometric quality. This last point is also reinforced by the usage of only sources flagged with high photometric quality in the 2MASS catalogs. The calibration zero point is calculated and applied individually for each VVV tile catalog. The final J, H, Ks magnitudes in the calibrated catalogs are in the usual 2MASS system.

\subsection{Completeness limit} \label{completeness}

Due to the different pixel size of the different surveys (GLIMPSE-II, 2MASS, VVV), completeness limits changes as a function of wavelength and depends on the field direction.
We divided the GLIMPSE II fields to subfields of  15' $\times$ 15' each to ensure a  sufficient number of stars (in the order of at least 200)  for our fitting process (see Sect.~\ref{method}). The completeness limit for every sub-field  was then calculated in each filter. In order to estimate the completeness we use star counts as a function of magnitude where the peak of each histogram in each filter  gives the corresponding completeness limit. The catalogue becomes severely incomplete beyond the point where the star counts reach maximum.
  We are aware that this is  a simple treatment of the completeness limit. However, we see here also clearly how strongly the completeness limit changes along different lines of sight as demonstrated \citep{saito2012} using artificial sources. 
  
 Fig.~\ref{fig1} shows the star counts as a function of magnitude for GLIMPSE-II and 2MASS catalogues for two subfields, one located at (l,b)=(0,+1) which is our reference field (the so-called C32 field) due to its very homogeneous and low extinction \citep{omont1999, omont2003}. The second field is located at  (l,b)=(0,-1)  selected here for its  higher and more clumpy extinction \citep{ojha2003}. Due to crowding and extinction the completeness limit
can change within one or two magnitudes in the infrared bands. Clearly seen is also the difference in the completeness limit at [3.6] and [4.5] compared to [5.8] and [8.0]
for both fields indicating the higher sensitivity of [3.6] and [4.5].

Figure~\ref{fig2} shows the same fields this time with star counts histograms from the  VVV data. Note the ``artificial'' raise at $\rm Ks=12^m$ which is the upper brightness limit of VVV. 
We decide to use in the following analysis the VVV data only from $\rm Ks > 12^m$. Figure ~\ref{fig2} demonstrates a clear gain in  sensitivity of VVV in comparison with 2MASS near-IR photometry. As shown by Saito et al. (2012), the VVV data is confusion limited. However, due to not fully matched 
spatial scales of the two catalogues and the fact that the overlap between 2MASS and VVV is in the magnitude range where the completeness between the two catalogues is very different, we do not investigate this issue further (this is beyond the scope of this paper), but it should be kept in mind when assessing the results.

We divided the VVV data in the same sub-fields as in  the GLIMPSE-II data set.

\section{The Besan\c{c}on model}

\subsection{Main description}

The Besan\c{c}on Galaxy model is a stellar population synthesis model of the Galaxy based on the scenario of formation and evolution of the Milky Way  \citep{robin2003}. It simulates the stellar contents of the Galaxy by using the four distinct stellar populations: the thin disc, the thick disc, the bulge and the spheroid.  It also takes into account the dark halo and
a diffuse component of the interstellar medium. For each population, a star formation-rate history and initial mass function are assumed which allows
to generate stellar catalogs for any given direction, and returns for each simulated  star its magnitude, color, and distance as well as kinematics and other stellar parameters. The model, in its present form,   can simulate  observations in many  combinations of different photometric systems from the UV to the Infrared. The photometry is computed using stellar atmosphere models (Basel 3.1) from \citet{lejeune1997} and \citet{lejeune1998}. However, while the Basel 3.1 stellar library is reliable for temperatures higher than 4000\,K,  atmospheric models become much more complex due to the appearance of molecules such as TiO, VO, $\rm H_{2}O$ for cooler stars. As we have a significant contribution of M giants in our study we decided to adapt a more realistic $\rm T_{eff}$ vs. color relation extending to cooler objects of about 2500\,K. This new relation is described in Sect.~\ref{input}.

Recently, a new Bulge model has been proposed by \citet{robin2012} as the sum of two ellipsoids: a main component that is the  standard boxy bulge/bar, which dominates the counts up to latitudes of about 5$\degr$, and a second ellipsoid that has a thicker structure seen mainly at higher latitudes.
The resulting 2MASS color-magnitude diagrams as well as star counts of this two component model clearly indicate the large improvement (see \citealt{robin2012}). \citet{uttenthaler2012} show also an excellent agreement in the kinematics based on radial velocity measurements.

For  more detailed information about the Besan\c{c}on model  we refer to \citet{robin2003} and its most recent update in \citet{robin2012}.

\begin{figure}[htbp]
   \includegraphics[width=9cm]{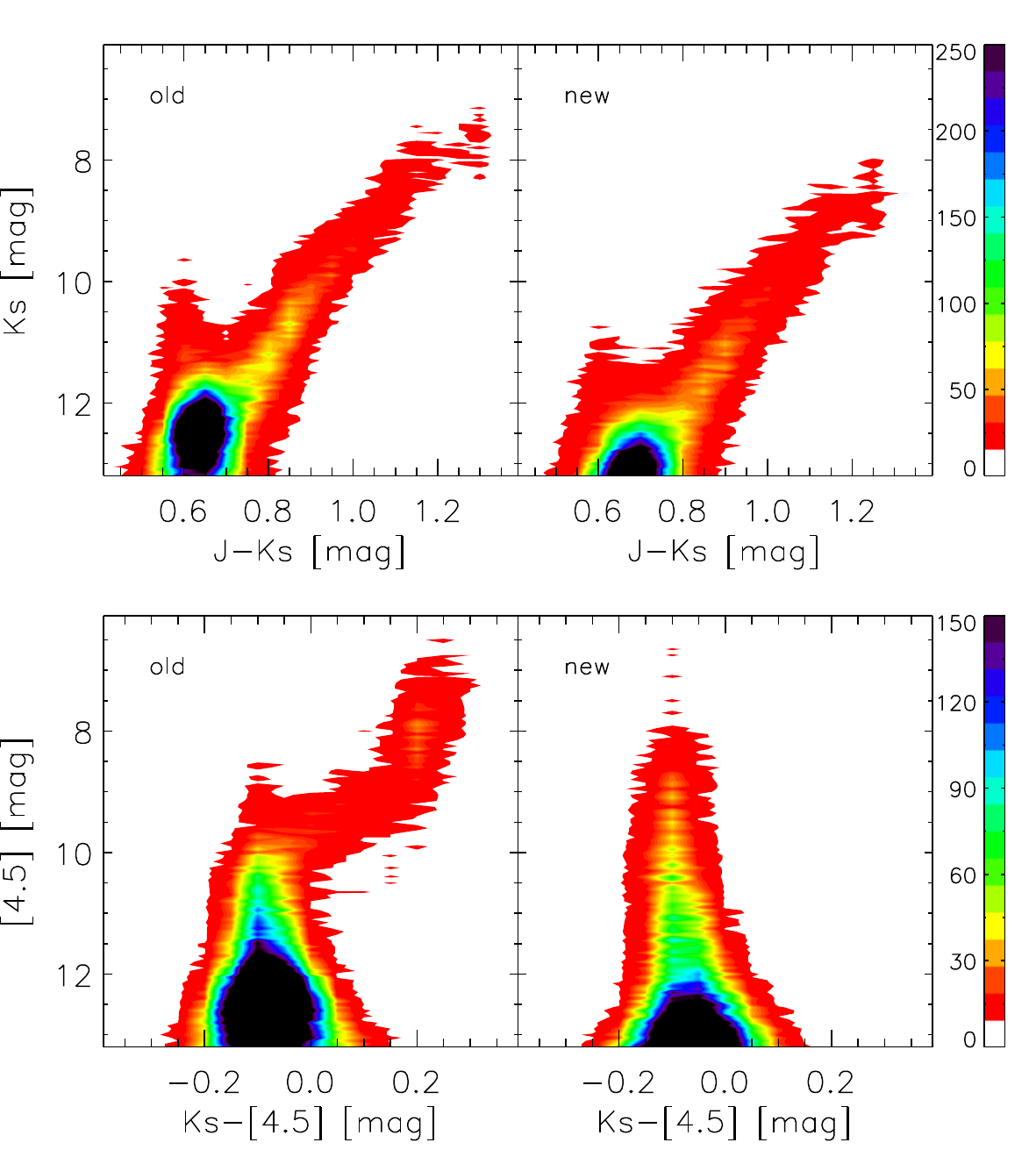}
\caption{Comparison of the the simulated color-magnitude diagrams using the old $\rm T_{eff}$ vs. color relation (left) compared to the new one (right) for the reference field (l=0.00 and b=1.00). The colorbar indicates the number of the stars.}
\label{fig3a}
\end{figure}

\begin{figure}[htbp]
   \includegraphics[width=9cm]{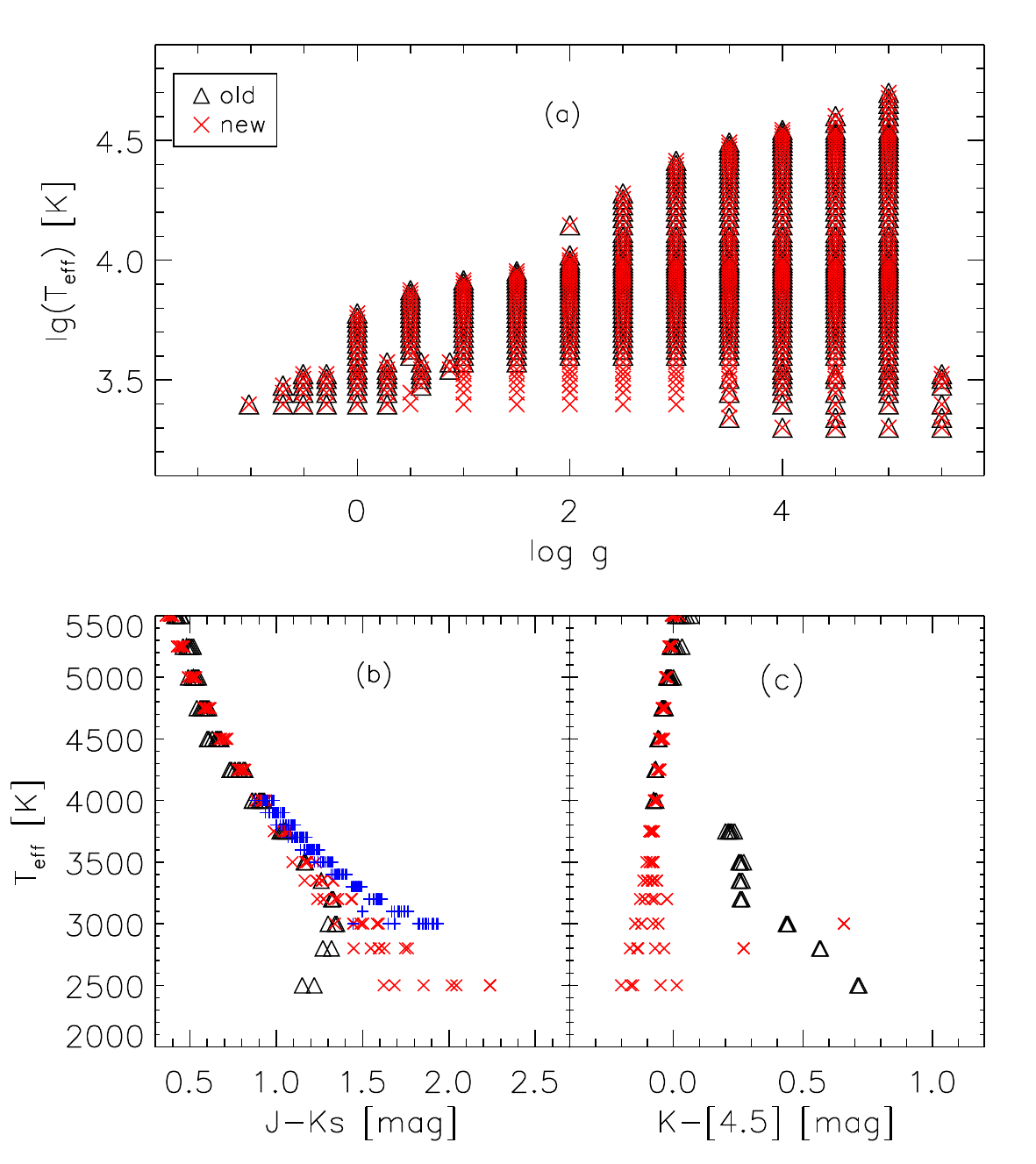}
\caption{(a): log\, $\rm T_{eff}$ vs log\,g grid in the Besan\c{c}on model. Black triangles indicate the coverage of the Besan\c{c}on model (Robin et al. 2003) while  red stars show the new extended grid. (b):   J--Ks vs. $\rm T_{eff}$ relation. Black triangles show the old relation while red crosses the new one. The blue symbols  is the relation for M giants from \citet{houdashelt2000}. (c). Similar as (b) but for K--[4.5] vs. $\rm T_{eff}$.}
\label{fig3b}
\end{figure}

\begin{table*}[htbp!]
\caption{Extinction as a function of Galactic longitude and latitude as well as distance based on VVV data.}
\centering
\tabcolsep 5.8pt
\begin{tabular}{cccccccc}
\hline\hline                        
l	&	b	&$E(J-Ks)_{1-10\,kpc}$&$E(H-Ks)_{1-10\,kpc}$&$\sigma E(J-Ks)_{1-10\,kpc}$&$\sigma E(H-Ks)_{1-10\,kpc}$ \\
\hline

\end{tabular}
\tablefoot{For each position we give the E(J--Ks),  E(H--Ks) as well as the corresponding sigma for each distance bin starting from 1 to 10\,kpc. }
\end{table*}

\begin{table*}[htbp!]
\caption{Extinction as a function of Galactic longitude and latitude as well as distance based on GLIMPSE-II data. }
\centering

\tabcolsep 5.8pt

\scalebox{0.71}{\begin{tabular}{@{}lllllllllll@{}}

\hline\hline                        

l	&	b	& $E(Ks-[3.6])_{1-10\,kpc}$&$E(Ks-[4.5])_{1-10\,kpc}$&$E(Ks-[5.8])_{1-10\,kpc}$&$E(Ks-[8.0])_{1-10\,kpc}$&$\sigma E(K-[3.6])_{1-10\,kpc}$&$\sigma E(Ks-[4.5])_{1-10\,kpc}$&$\sigma E(Ks-[5.8])_{1-10\,kpc}$&$\sigma E(Ks-[8.0])_{1-10\,kpc}$\\

\hline

\end{tabular}
}
\tablefoot{For each position we give the E(Ks--[3.6]),  E(Ks--[4.5]), E(Ks--[5.8]), E(Ks--[8.0]) as well as the corresponding sigma for each distance bin starting from 1 to 10\,kpc.}

\end{table*}

\subsection{Implementation of IRAC colors and   a new  temperature-color relation} \label{input}

Reliable temperature-color relations are of extreme importance for comparison of  the synthetic color-magnitude and color-color diagrams with the observed ones.  We used the reference field at (l,b)=(0,+1) to check the new synthetic colors in the [IRAC] filters. Fig.~\ref{fig3a} shows a kink in the $\rm Ks-[4.5]$ vs. [4.5] color magnitude diagram at $\rm  [Ks]-[4.5]\sim+0.1,[4.5]\sim10$ which we identify clearly also in  the corresponding  temperature  vs. color relation (see Fig.~\ref{fig3b}c). While the Basel3.1 library predicts for $\rm T_{eff} < 3000\,K$
 decreasing J--Ks values, our new adopted relation shows a steady increase in J--Ks until $\rm T_{eff} = 2500\,K$. We also noticed that the full parameter space in the $\rm log\,T_{eff}$ vs. log g plane was not covered in the previous model, especially there is a gap in the temperature range between 2500--4000\,K for $\rm 0.5 < log\,g < 3$.

We updated the temperature-color relation and filled in the gaps, by using the isochrones from \citet{girardi2010}. These isochrones include the TP-AGB phase \citep{marigo2008} including mass-loss. Starting from a group of isochrones with ages ranging between 5--10 Gyr we derived a new temperature vs. color relation for metallicities between [Fe/H]=-2 and [Fe/H]=0.0. These ages and metallicities are characteristic of the Bulge populations (e.g. \citealt{zoccali2003}, \citealt{zoccali2008}, \citealt{bensby2011}). The comparison between the color-temperature relations from the Besan\c{c}on model (so called "old" relation) and the one derived by us (so called "new" relation) is shown in Fig.~\ref{fig3b}b and c. For comparison we show
the J--Ks  vs.  $\rm T_{eff}$ relation obtained by \citet{houdashelt2000} for M giants.  The clear  difference between the new and old temperature vs. color relations for $\rm T_{eff} < 4000\,K$ is obviously a culprit for the bent in the color-magnitude diagram (Fig.~\ref{fig3a}), which disappears after adopting our new relation. We also see that using the new relation the J--Ks colors will be systematically redder by  about 0.05 mag.

\begin{figure*}[htbp!]
   \includegraphics[width=6cm]{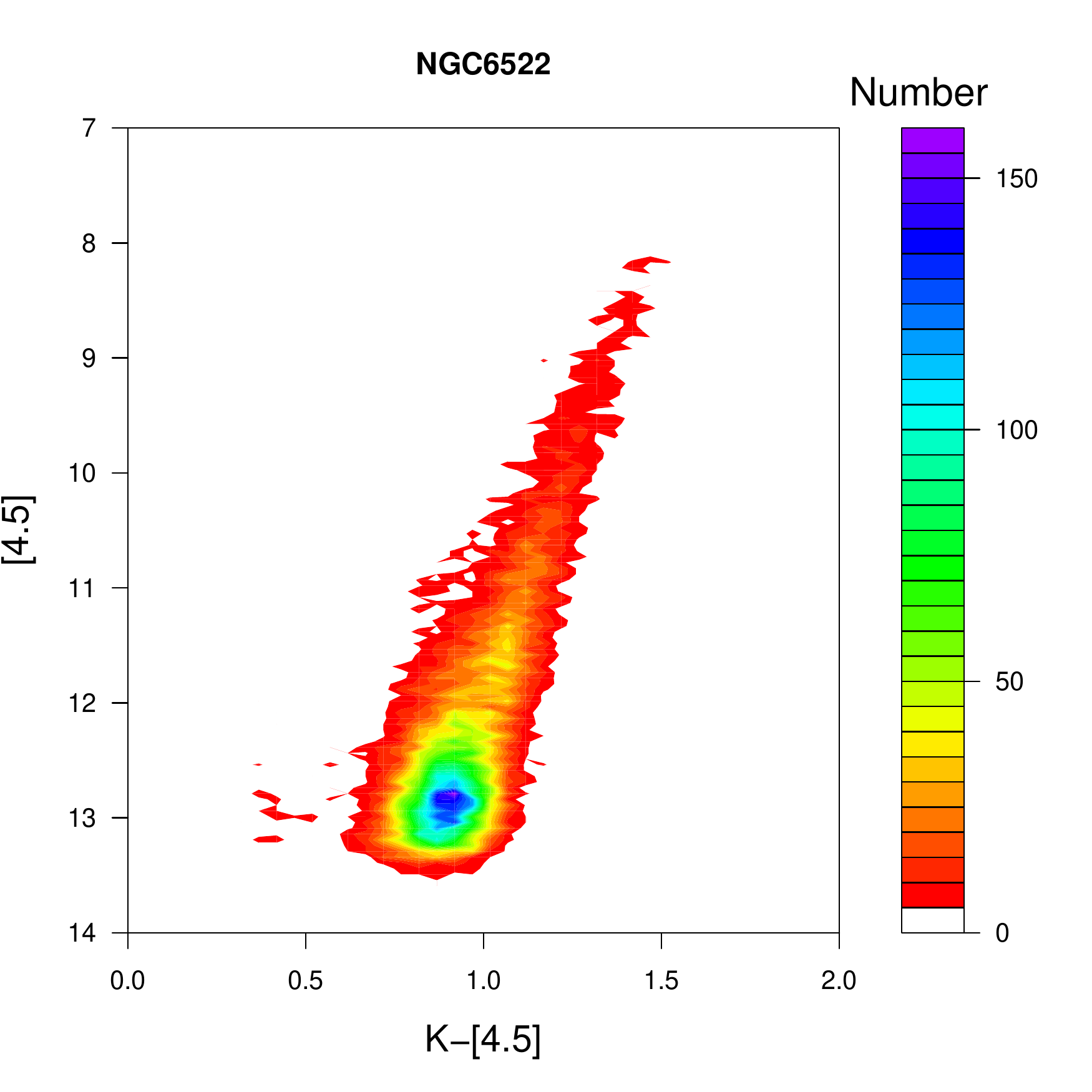}  \includegraphics[width=6cm]{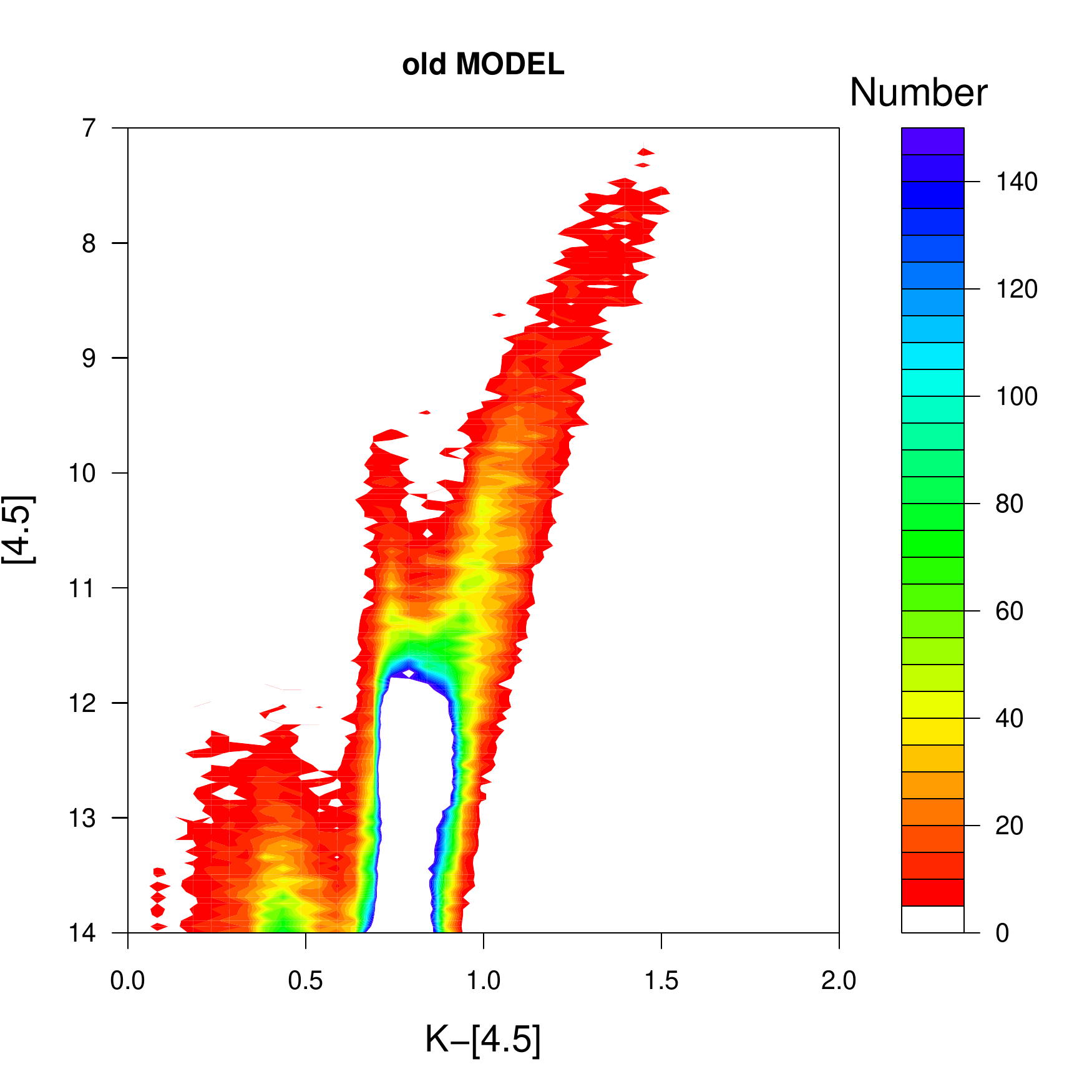} \includegraphics[width=6cm]{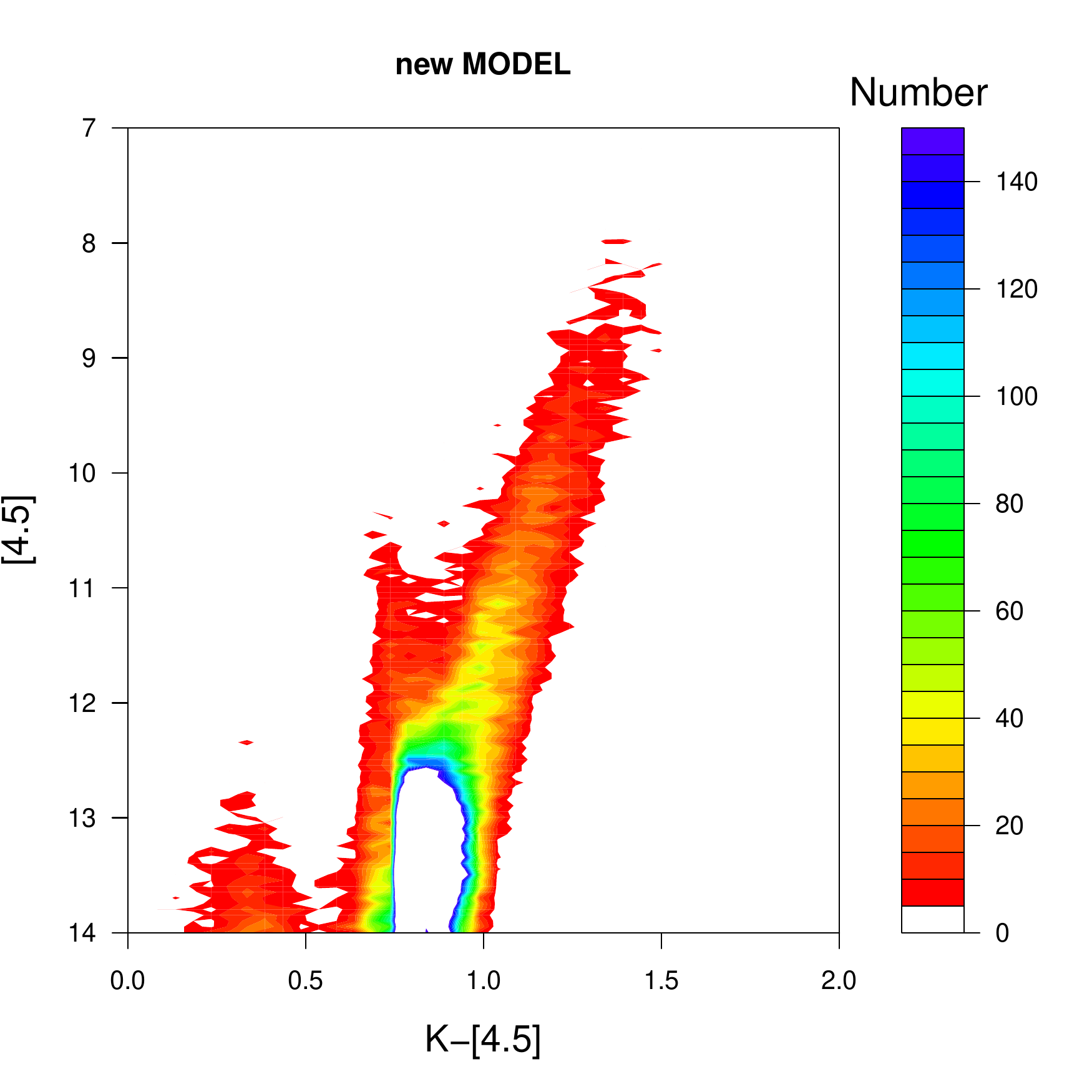}
\caption{Comparison of the Ks--[4.5] vs. [4.5] diagram of Baade's window (left panel) with the ``old'' (middle panel) and the ``new'' temperature vs. color relation (left panel). The data are from \citet{uttenthaler2010}. The colorbar indicates the number of the stars. }
\label{stefan}
\end{figure*}

 Figure \ref{stefan} shows the comparison of a color-magnitude diagram in a low-extinction field -the so-called Baade's window. We use here the catalog of \citet{uttenthaler2010}. One notices clearly that the predicted color in the ``old'' model is too blue compared to the observations and we see in addition a  much larger fraction of foreground objects ($\rm Ks-[4.5] < 0.5$) which we don't find in the data.

\section{Method}  \label{method}

We obtain the extinction by using the color excess between the intrinsic color (derived from the model defined as $\rm Co_{ins}$) and the observational apparent color
(derived from the observation, here we defined as $\rm Co_{obs}$). Assuming an extinction law, we get:
\begin{equation}
 A_\lambda = c_\lambda \times (\bar{Co_{obs}}- \bar{Co_{ins}})
\end{equation}
where $c_\lambda$ is related to the extinction coefficient. As we trace our fields close to the Galactic Center,  we used the extinction coefficients from \citet{nishiyama2009}. We ignore  here the effect of the broadband extinctions and the reddenings on the SED of the individual sources as our method does not allow for a more sophisticated treatment which could imply systematic errors.  This results in:\\

$ A_{Ks,J-Ks} = 0.528 \times ((J-Ks)_{obs}-(J-Ks)_{ins}) $

$ A_{Ks,H-Ks} = 1.61 \times  ((H-Ks)_{obs}-(H-Ks)_{ins}) $

$ A_{[3.6],Ks-[3.6]} = 1.005 \times ((Ks-[3.6])_{obs}-(Ks-[3.6])_{ins}) $

$ A_{[4.5],Ks-[4.5]} = 0.640 \times ((Ks-[4.5])_{obs}-(Ks-[4.5])_{ins}) $

$ A_{[5.8],Ks-[5.8]} = 0.562 \times ((Ks-[5.8])_{obs}-(Ks-[5.8])_{ins}) $

$ A_{[8.0],Ks-[8.0]} = 0.748 \times ((Ks-[8.0])_{obs}-(Ks-[8.0])_{ins}) $\\

\smallskip

\begin{figure}[htbp!]
   \includegraphics[width=9cm]{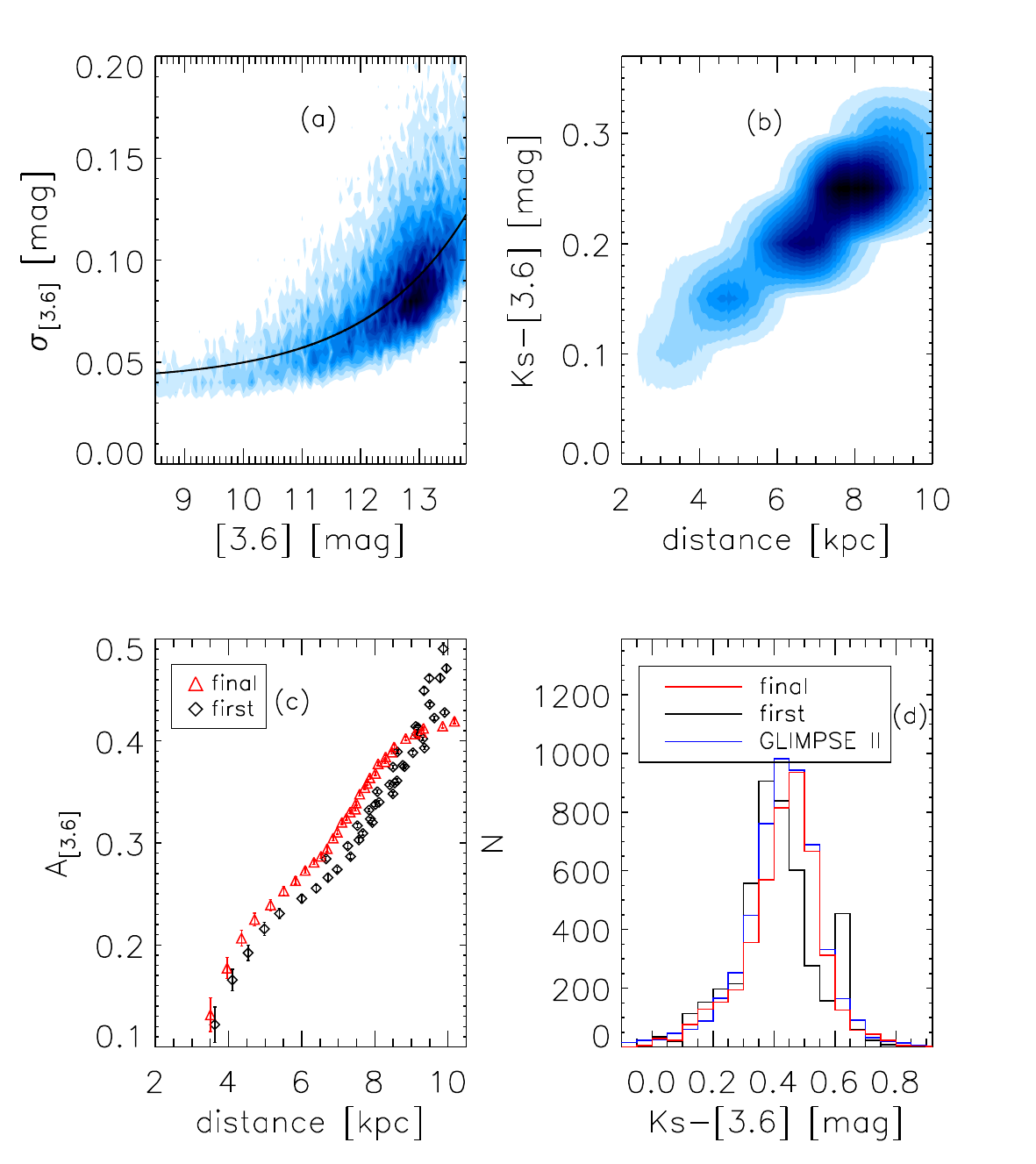}
\caption{The process to calculate the extinction of the example subfield (l,b) of (0.0,+1.0). From upper left to lower right, they are:(a) the errors of [3.6] vs. the [3.6] diagram, (b) the color Ks--[3.6] vs. distance diagram, (c) the  extinction at [3.6]  vs. the distance diagram where we show the first initial guess of extinction after the first iteration (black) and the final iteration after 20 loops (red) and (d)  the color Ks--[3.6] distribution of the data compared to  the first guess iteration (black) and the final one (red).}
\label{fig4}
\end{figure}

In the following we apply the same method as \citet{marshall2006} assuming that the distance grows as the apparent color/extinction increases. Fig.~\ref{fig4}b,c  shows an example of this relation for our reference field. Contrary to \citet{marshall2006} we use the full information of the stellar populations in the CMD, e.g. we include also the M dwarf population. While the local  dwarf population (d $<$ 1--2\,kpc) can be easily identified (see \citealt{marshall2006}) by simple photometric color criteria (e.g. J--Ks), this is not the case for dwarfs located at larger distances where they cannot be separated from giants. This is especially true for  the VVV data  where we are able to detect  dwarf stars up to larger distances. However, the model simulations show that the contribution of dwarf stars within the VVV completeness limit used in our study  is rather small ($\sim 1\,\%$)  but can increase up to 20-30\,\% going to fainter magnitudes beyond the VVV completeness limit.

The method can be described in following steps:

\begin{enumerate}

\item

First we add photometric errors and a diffuse extinction to the intrinsic colors $Co_{ins}$ from the model for each subfield. The assumption of a diffuse extinction is necessary to ensure that the colors of the model increase with distance. We assume a diffuse extinction of 0.7 mag/kpc in the V band. 
 We use exponential errors for the GLIMPSE-II data in all seven filters. Fig.~\ref{fig4}a shows an example of the photometric errors as a function of the [3.6] magnitude for our reference field (l,b) = (0.00, +1.00). For the VVV data, the error is much smaller and we adopted  a constant  error of 0.05\,mag \citep{saito2012}. This procedure provides a first set of simulated colors ($\rm Co_{sim,ini}$)

\item

Colors are then sorted, both for observed and simulated data ($\rm Co_{obs}$ and $\rm Co_{sim,ini}$), and normalized by the corresponding number of colors in each bin given by  $n_{bin}=floor(min([N_{obs},N_{bes}])/100)$.  $N_{obs}$ and $N_{bes}$ are the total number of observed stars (in each subfield) and of stars in the Besan\c{c}on model, respectively. This ensures that we have at least 100 stars per bin.

\item

In each corresponding color bin, the $Co_{ins}$ of the stars from the model data and the $Co_{obs}$  from the observational data are used to calculate the extinction using Eq.~1, as well as the median distance from the model. A first distance and extinction estimate   can be thus obtained ($A_{1st}$). Due to the saturation limit of VVV ($\rm Ks = 12$), we do not get enough stars in the first distance bins (0--3\,kpc). Therefore our 3D map starts at 4\,kpc for VVV, while at 3\,kpc for GLIMPSE-II data.

\item

The extinction  as well as the distance estimated from the third step is then directly applied to the intrinsic magnitudes in the model which gives us a  new simulated color  ($Co_{sim,1}=Co_{ins}+A_{1st}$).

\item
We construct  histograms for each color of the new simulated data ($Co_{sim,1}$) and the observational data ($Co_{ins}$)  using the same binsize (0.05\,mag).
The  $\chi ^2$ statistics \citep[][]{press1992,marshall2006} are then used to evaluate the similarity of both histograms, given by:

\begin{equation}
 \chi^2 = \sum _j {(\sqrt{N_{obs}/N_{sim}}n_{sim_j}-\sqrt{N_{sim}/N_{obs}}n_{obs_j})^2/(n_{sim_j}+n_{obs_j})}
\end{equation}

 where $N_{obs}$ and $N_{sim}$ are the total number of observed and simulated stars in each subfield, while 
$n_{obs_j}$  and $n_{sim_j}$  show the number of stars for the $j$th  color bin  of the observations and simulated data, respectively.

With the first extinction estimate applied we have now a new set of simulated data. For this new set of data, some of the stars may be fainter to fall outside the completeness limit while others of them may be brighter and come inside the limit. We go back to the second step with the new $\rm Co_{sim,1}$ which would be used iteratively to refine our results. In total we perform 20 iterations for each sub-field and each color. The minimum $\chi ^2$ is then our final result. We have chosen 20 loops to ensure that we always find a minimum value. In most of the cases the $\chi ^2$ converges  after one or two loops to a minimum  and then becomes stable in the following loops.

\end{enumerate}

Fig. ~\ref{fig4}c shows an example of the iteration process using the Ks--[3.6] color of the GLIMPSE-II data in our reference field (l,b)=(0.00, 1.00). We see clearly the difference between the first extinction estimate and our final extinction determination after 20 loops. Note that the we adapt  the  same completeness limit as in the observations for the model simulations (see Sect.\ref{completeness}).

\begin{figure}[htbp!]
   \includegraphics[width=9cm]{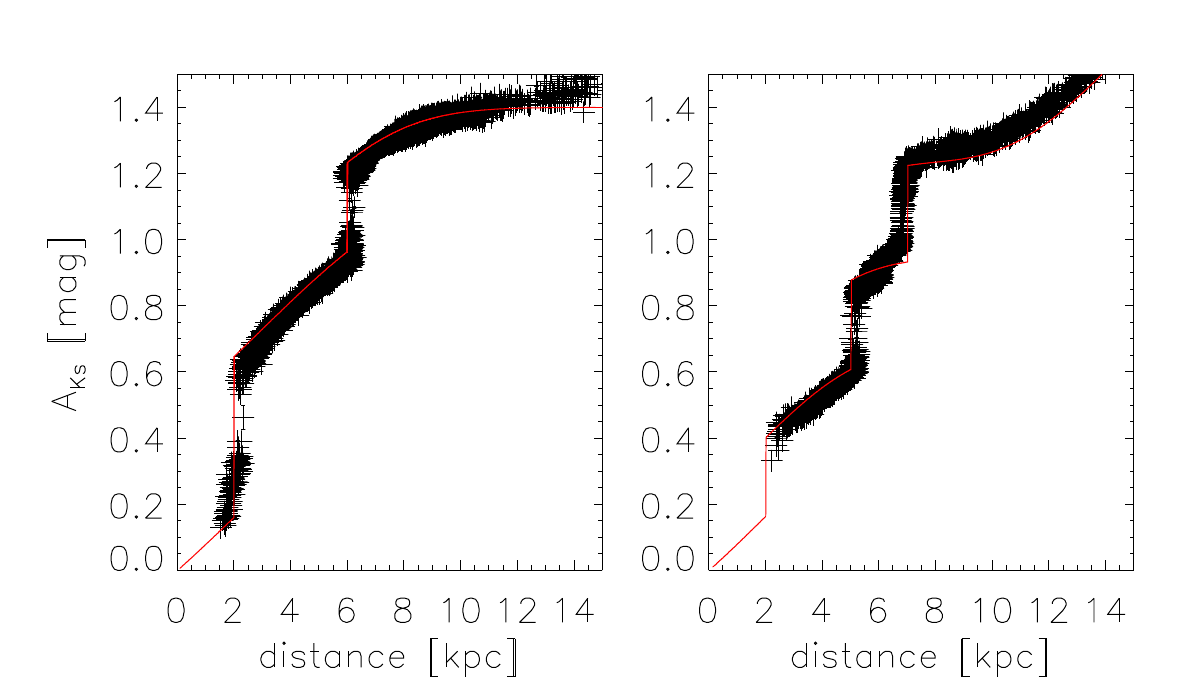}
\caption{The test of our method. The red line is the artificial extinction with clouds. The black crosses show our derived extinction values. The left panel is of field (l,b)=(0.00,+1.00) and the right (l,b)=(5.00,0.00).}
\label{testhks}
\end{figure}

To test our method, we used artificial data by adding manually extinction values to the intrinsic magnitude of the  model. We used different lines of sight with a different number of clouds. Figure \ref{testhks}  shows two examples of several clouds. For field (l,b) at (0.00, 1.00), one located at a  distance of 2\,kpc with  $\rm A_{V} = 4.5\,mag$, the second one at 6 \,kpc with $\rm A_V = 2.5\,mag$.  For the field (l,b)=(5.00,0.00), we assumed three clouds: $\rm A_{V}= 2.20$ \,mag at 2.0 \,kpc, 2.50 \,mag at 5.0 \,kpc and 2.70 \,mag at 7.0 \,kpc. We applied our method to our simulated data. We got 328.000 sources for the first field and 182.000 for the second field. We cut the data at $\rm Ks= 18\,mag$ artificially, which goes much deeper than the observational data. Figure \ref{testhks} shows that our method successfully derives the extinction along the distance and that we are able to trace this multi-cloud structure.
Modifying the input parameters of the galactic model such as galactic scale length, the size of the hole of the galactic disc  or changing the Bulge luminosity function  could lead to systematic differences in our derived extinction map. But this systematics, as shown in Marshall et al. (2006), is of the order of the random uncertainty of the method. 

\begin{figure*}[htbp!]
   \includegraphics[width=15cm]{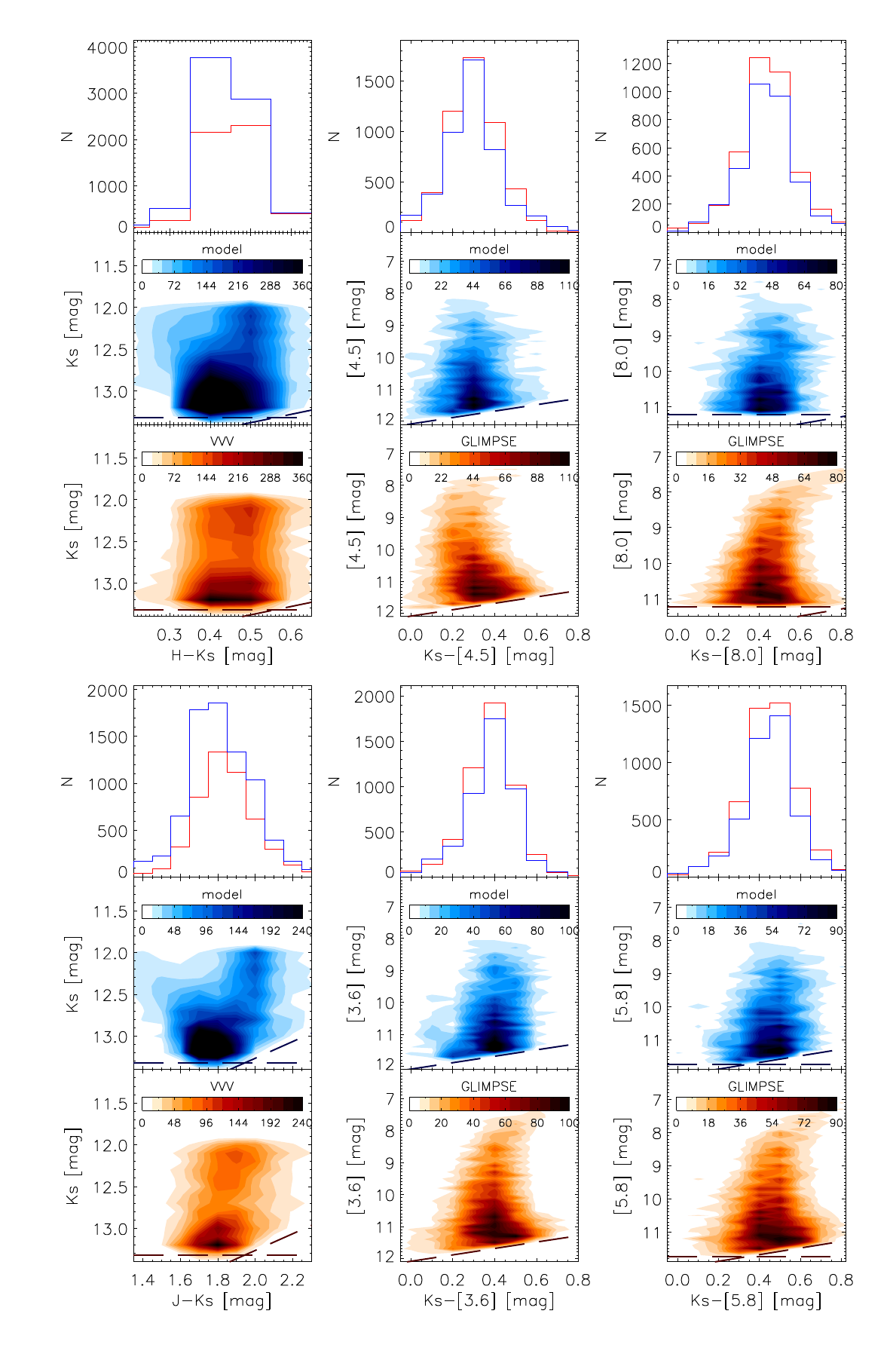}
\caption{CMD and color distribution for all the colors of the field l=0.00, b=1.00. The lower panel shows the observed CMD, the middle panel the synthetic CMD with the best-fitted extinction and the upper panel the histograms in the color distribution. The red color scale (lower pannel) shows the observations (noted as 'VVV or GLIMPSE' in the title of the inner colorbar) while the blue for simulated data (noted as 'model'). The dashed lines show the completeness limit. The colorbar inside the CMD shows the number of stars of each corresponding dataset.}
\label{fig7}
\end{figure*}

\section{Results and discussion}

The full results are listed in Table 1 and Table 2, both available in electronic forms at the CDS\footnote{Table 1 and Table 2 are only available in electronic form
at the CDS via anonymous ftp to cdsarc.u-strasbg.fr (130.79.128.5) or via http://cdsweb.u-strasbg.fr/cgi-bin/qcat?J/A+A/.}. Each row of Table 1 and Table 2 contains the information for one line of sight: Galactic coordinates along with the measured quantities for each distance bin E(J--Ks), E(H--Ks) and respective uncertainties (VVV, Table 1) as well as E(Ks--[3.6]), E(Ks--[4.5]), E(Ks--[5.8]), E(Ks--[8.0]) and their uncertainties (GLIMPSE-II, Table 2). These results will be also added into the BEAM calculator\footnote{http://mill.astro.puc.cl/BEAM/calculator.php}  webpage \citep{gonzalez2012} where the 2D VVV extinction map is already available to the community. Users of the BEAM calculator can choose to retrieve the extinction calculation adopting a specific reddening law and distance.

\begin{figure*}[htbp!]
   \includegraphics[width=15cm]{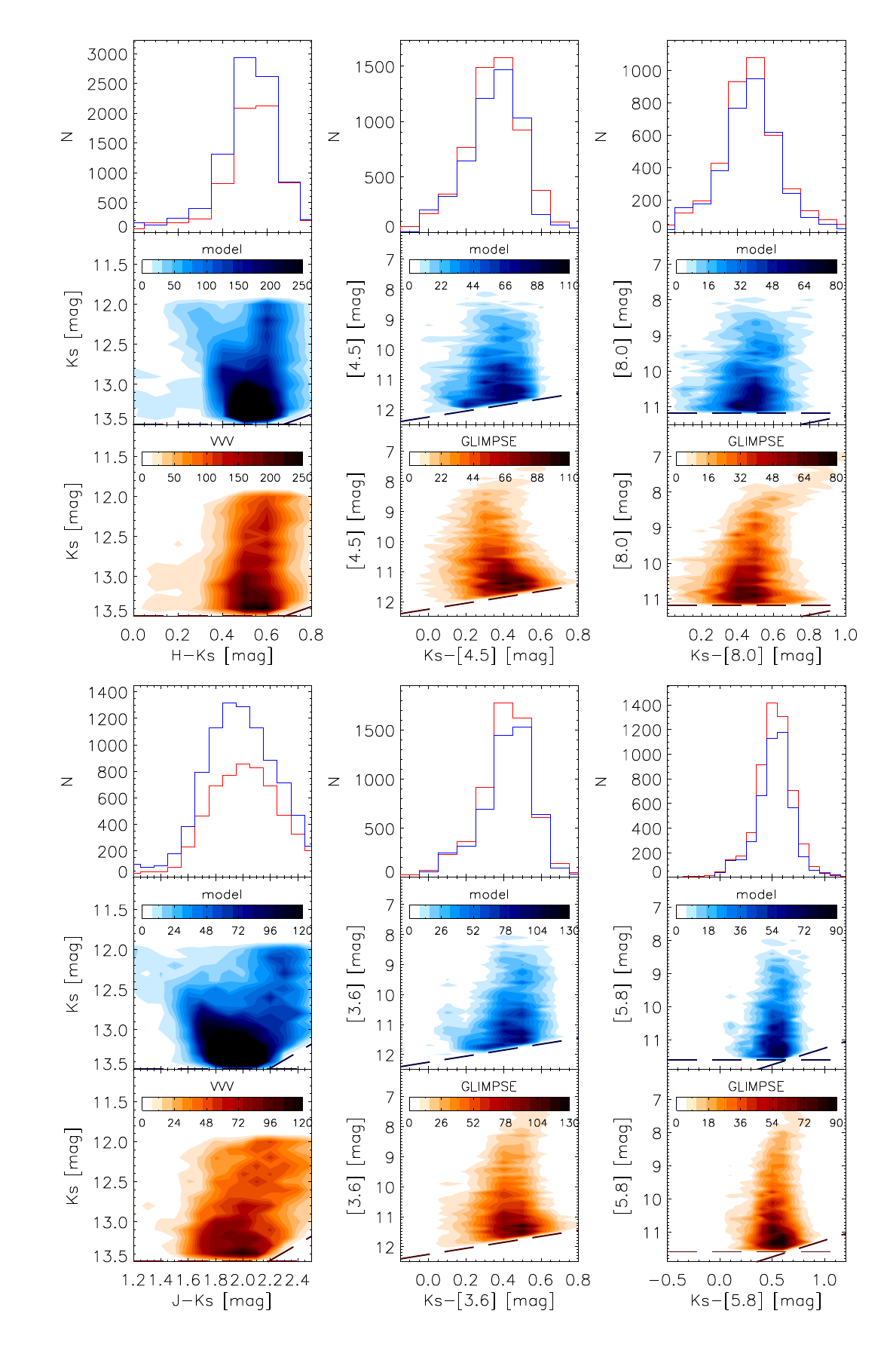}
\caption{CMD and color distribution for all the colors of the field l=0.00, b=-1.00, same as Fig.~\ref{fig7}.}
\label{fig8}
\end{figure*}

\subsection{Extinction along different lines of sight}

As an example of our results we discuss in more detail here two specific lines of sight: one the reference field (l,b)=(0.00,1.00) and the other is the field at (l,b)=(0.00,-1.00).

Figure~\ref{fig7}  and Figure~\ref{fig8} show the corresponding color-magnitude diagrams and color distribution for the observational data and the simulated data with the best-fitted final extinction in each filter.  Besides the general good agreement in the shape of the CMD and in the color distribution, we want to emphasize also  that the  number densities in the CMDs between the observations and the model agree very well. However we notice some  interesting features by comparing the CMDs between the model and the observations:

\begin{itemize}
\item We see in the [5.8] vs. Ks--[5.8] and especially in the [8.0] vs. Ks--[8.0] GLIMPSE-II diagram clearly the Asymptotic Giant Branch  phase. The tip of the AGB  is approximately at $\rm [8.0]$ a bit brighter than 8.0\,mag. This phase is not reproduced by the model and needs  to be improved.
\item Contrary to GLIMPSE-II, the model predicts too many stars for VVV in all filters at the fainter end of the CMD. We suspect that the completeness limit of  VVV  is underestimated. \citet{saito2012} did for two VVV fields a complete analysis of the completeness limit by using artificial sources. They found that the source detection efficiency reaches 50\% for stars with 16.4 $<$ Ks $<$  16.9 mag for the tile b314 located at (l,b)=(352.21,-0.92). Our derived completeness limit is Ks = 14.9 which reaches according to \citet{saito2012} the 75\% completeness level. A detailed analysis of the completeness limits of the VVV survey using artificial sources would be needed.
\item   We notice that the model has produced a second branch in the J--Ks vs Ks diagram at about Ks = 12.5\,mag which we do not find in the VVV data (see Fig.~\ref{fig8}). This branch is related mainly to the galactic disc K-giant population, which is clearly too prominent in the model. In  contrary to the  J--Ks and H--Ks color, the J--H color shows a large scatter in the  temperature vs. color relation.  This scatter is mainly due to the large influence of  J--H to metallicity and gravity (see also \citealt{bessell1989}).  We decided therefore not to use the J--H color in our analysis.

\end{itemize}

\begin{figure}[!htbp]

   \includegraphics[width=9cm]{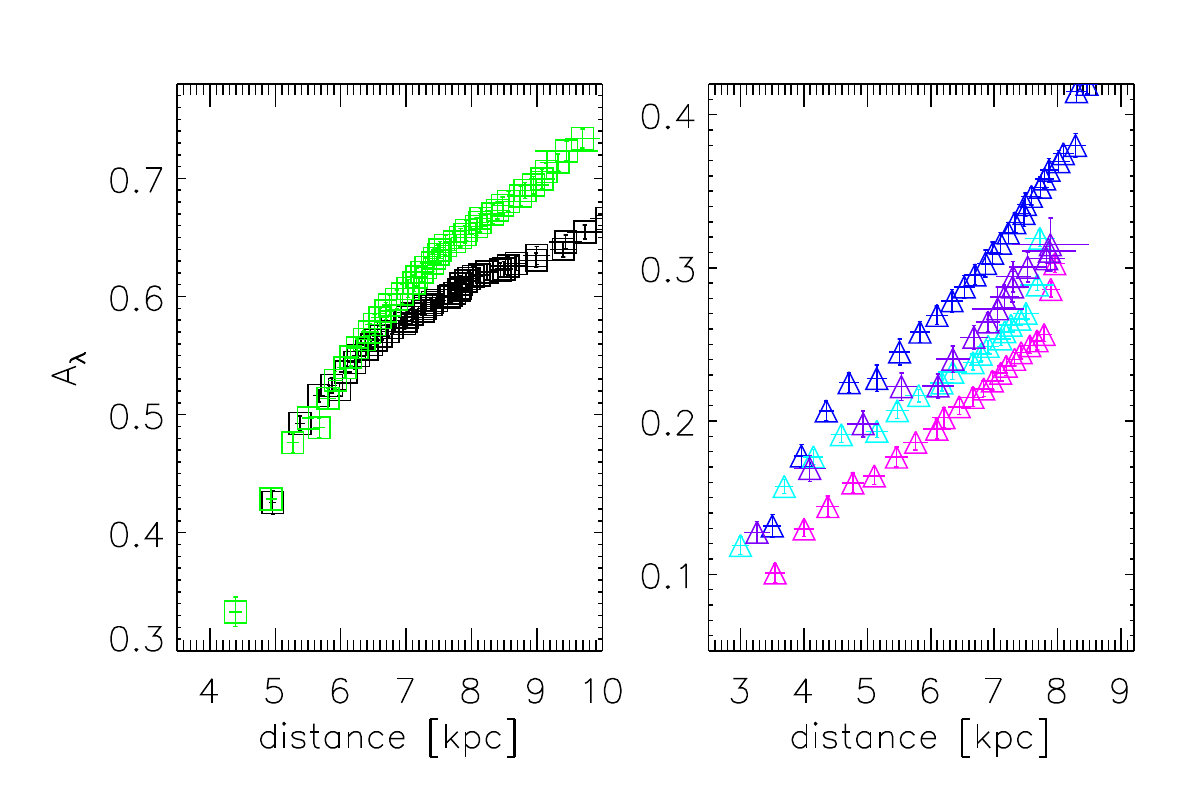}
\caption{The distance vs. extinction diagram for sub-field l=0.00, b=+1.00. Left panel: green color is $\rm A_{Ks,H-Ks}$ and black is$\rm A_{Ks,J-Ks}$; Right panel: blue color is $\rm A_{[3.6],K-[3.6]}$, violet color  $\rm  A_{[4.5],Ks-[4.5]}$,  light blue color $\rm A_{[5.8],K-[5.8]}$, magenta  $\rm A_{[8.0],K-[8.0]}$.}
\label{fig5a}
\end{figure}

\begin{figure}[!htbp]
   \includegraphics[width=9cm]{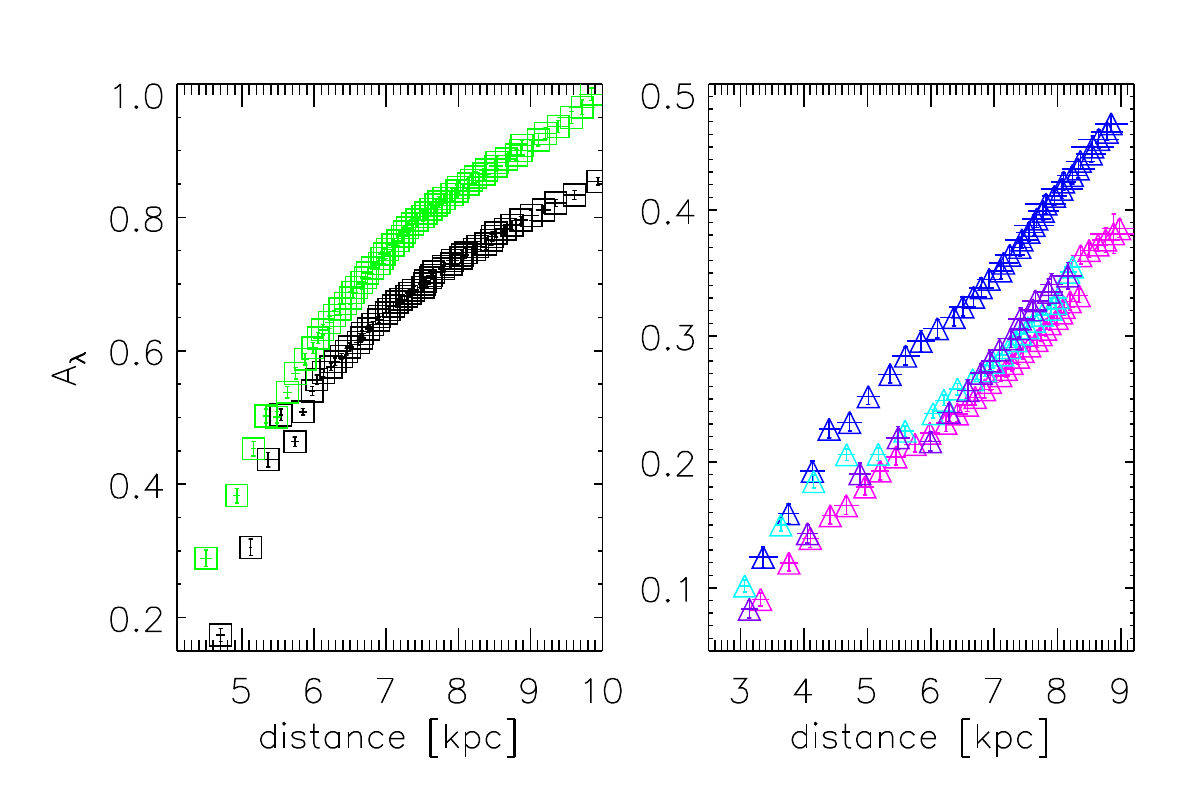}

\caption{The distance vs. extinction diagram for sub-field l=0.00, b=-1.00. The symbols are the same as in Fig. \ref{fig5a}.}
\label{fig5b}
\end{figure}

\begin{figure}[!htbp]
   \includegraphics[width=9cm]{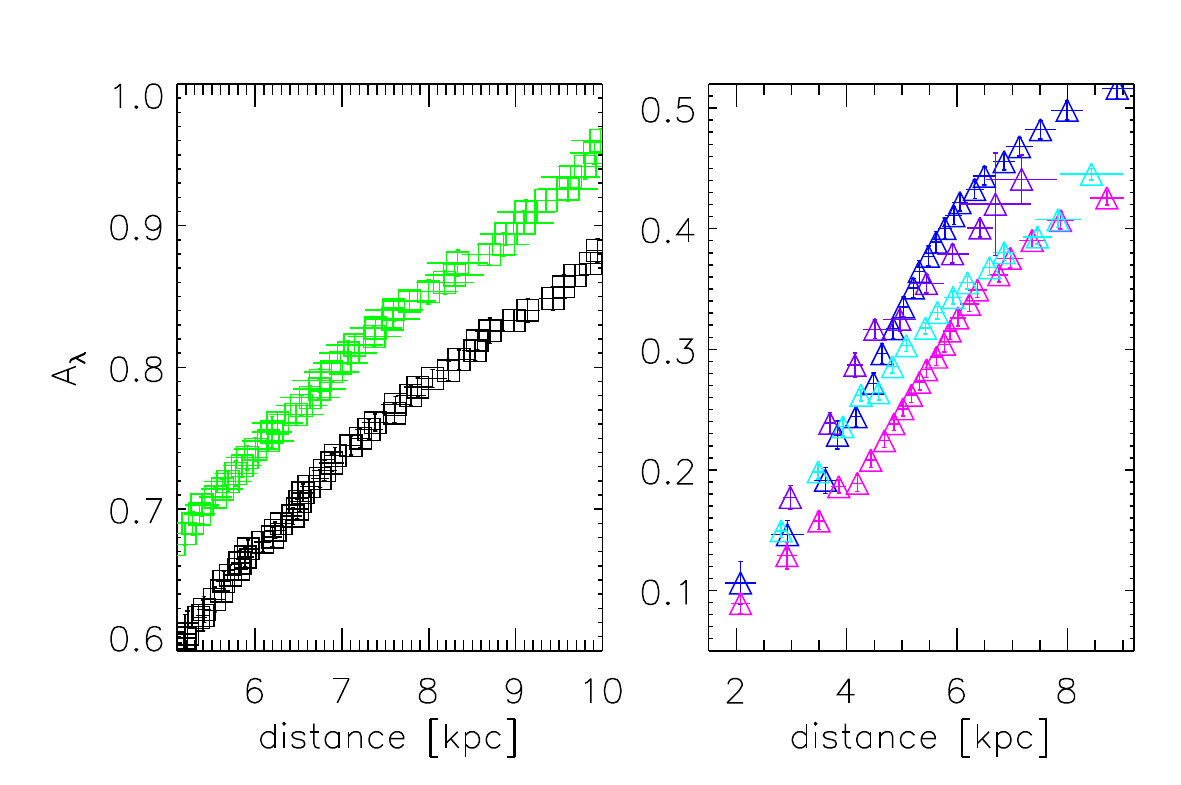}

\caption{The distance vs. extinction diagram for sub-field l=10.00, b=1.00. The symbols are the same as in Fig. \ref{fig5a}.}
\label{fig5c}
\end{figure}

\begin{figure}[!htbp]
   \includegraphics[width=9cm]{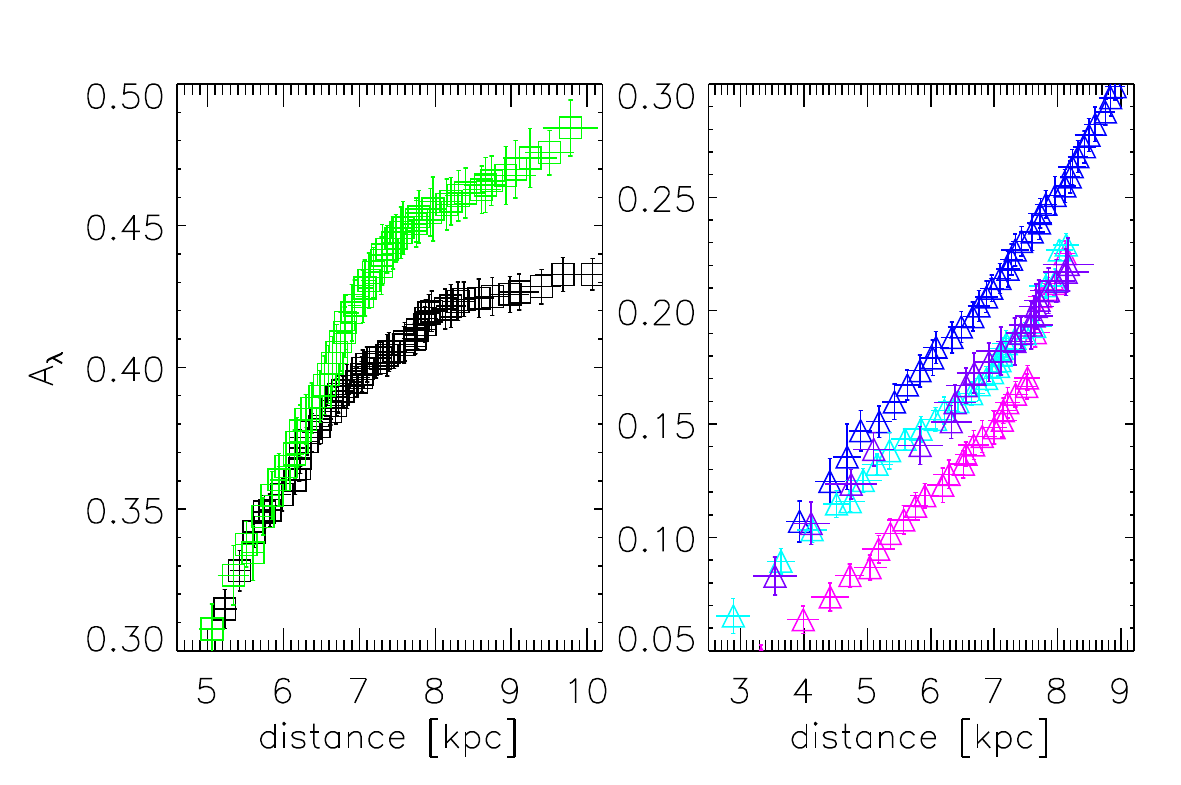}

\caption{The distance vs. extinction diagram for sub-field l=0.00, b=+1.75. The symbols are the same as in Fig. \ref{fig5a}.}
\label{fig5d}
\end{figure}

\begin{figure}[!htbp]
   \includegraphics[width=9.5cm]{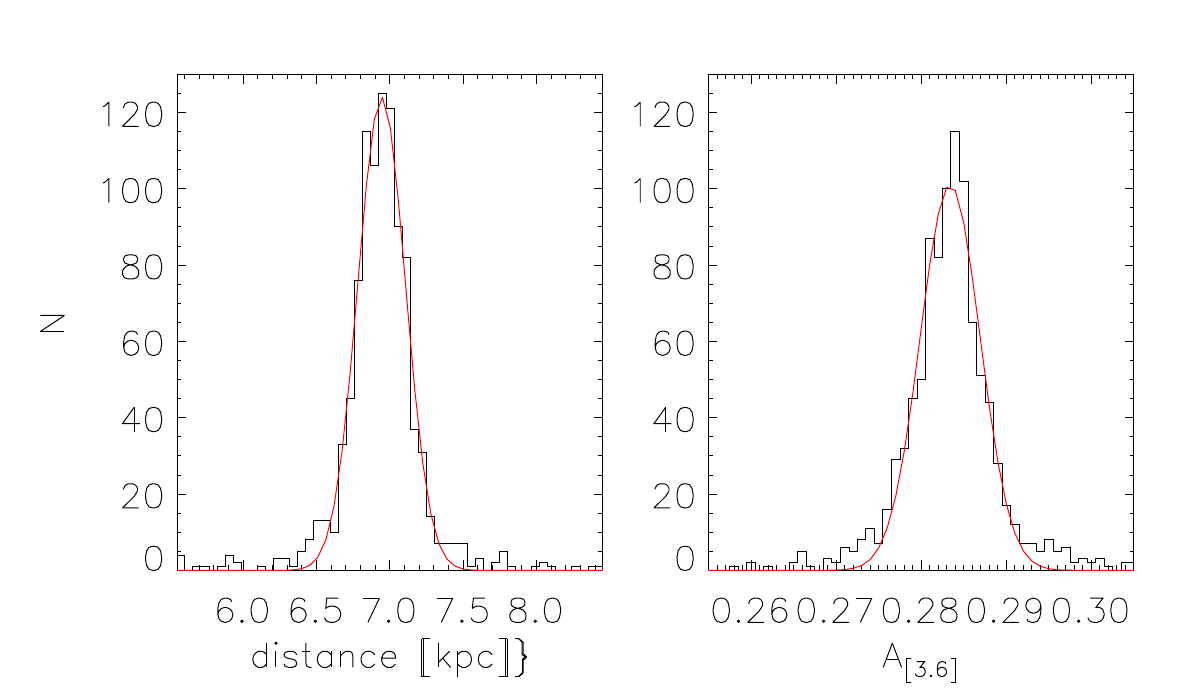}
\caption{The distribution of the distance (left) and the extinction (right) derived from 1000 samples with bootstrap method.}
\label{booterr}
\end{figure}

Figure~\ref{fig5a} to Figure~\ref{fig5d} show how interstellar extinction changes with distance in the different  bands for four subfields. Different symbols stand for different data set: squares for VVV data and triangles for GLIMPSE-II data while  different colors denote the various filters. The errors of the derived distances and the extinction were calculated using the bootstrap method \citep{wall2003}. We re-sampled the original datasets of the observational and the model data  1000 times. The r.m.s scatter gives  the uncertainty of our results.
 Fig.~\ref{booterr} shows a typical example.  It shows the distributions of our results obtained for the 1000 samples obtained for one colour bin (Step 3 in Sect. 4) for the GLIMPSE-II data in [3.6], left for the distance and right for the extinction. A Gaussian fit was made to the distributions, where the full-width-half maximum of the distribution gives the uncertainty in distance and extinction.  This process is applied for each bin and each filter.  The extinction changes along with the distance. As expected, it is higher for increasing distance.
As seen in  Fig.~\ref{fig5a} to Figure~\ref{fig5d}, extinction gets flatter beyond 7\,kpc. As shown by \citet{marshall2006} this is probably due to the fact that inside the molecular ring the density of interstellar matter is lower than outside. It has been shown also that this region contains the dust lanes of the bar (\citealt{marshall2009}), which can explain that at positive longitudes one can see an increase of the absorbing matter after this plateau, while it appears further away at negative longitudes.

\begin{figure*}[!htbp]
   \includegraphics[width=18cm]{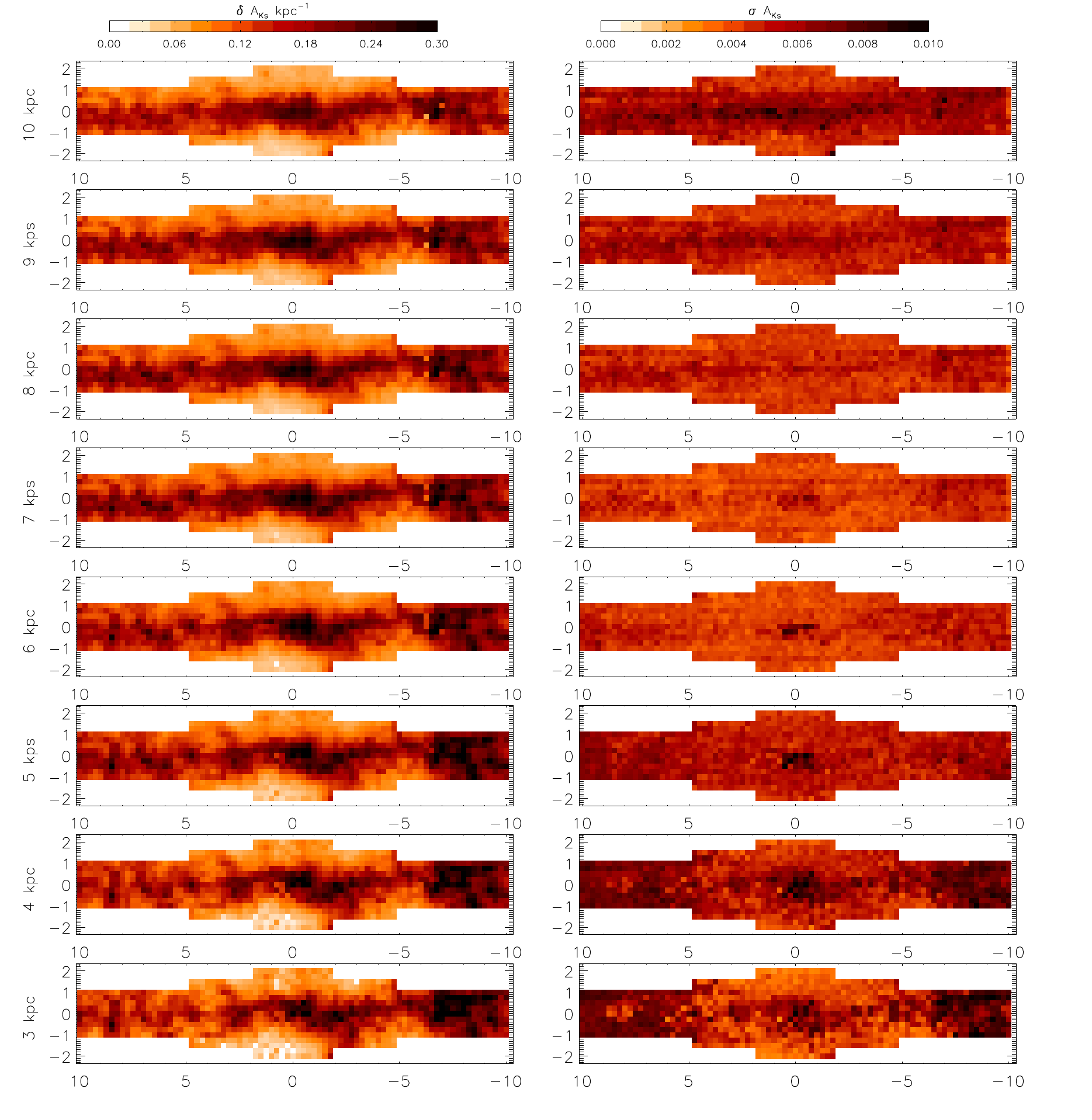}
\caption{3d extinction and error maps for $\rm A_{Ks}$. X-axis denote galactic longitude while y-axis galactic latitude. The units are in $\rm \delta A_{Ks}\,kpc^{-1}$. The map shows the weighted average extinction converted from the seven individual colors excesses using the reddening law derived from Sect.~\ref{coeff}.}
\label{fig9}
\end{figure*}

\begin{figure*}[!htbp]
   \includegraphics[width=18cm]{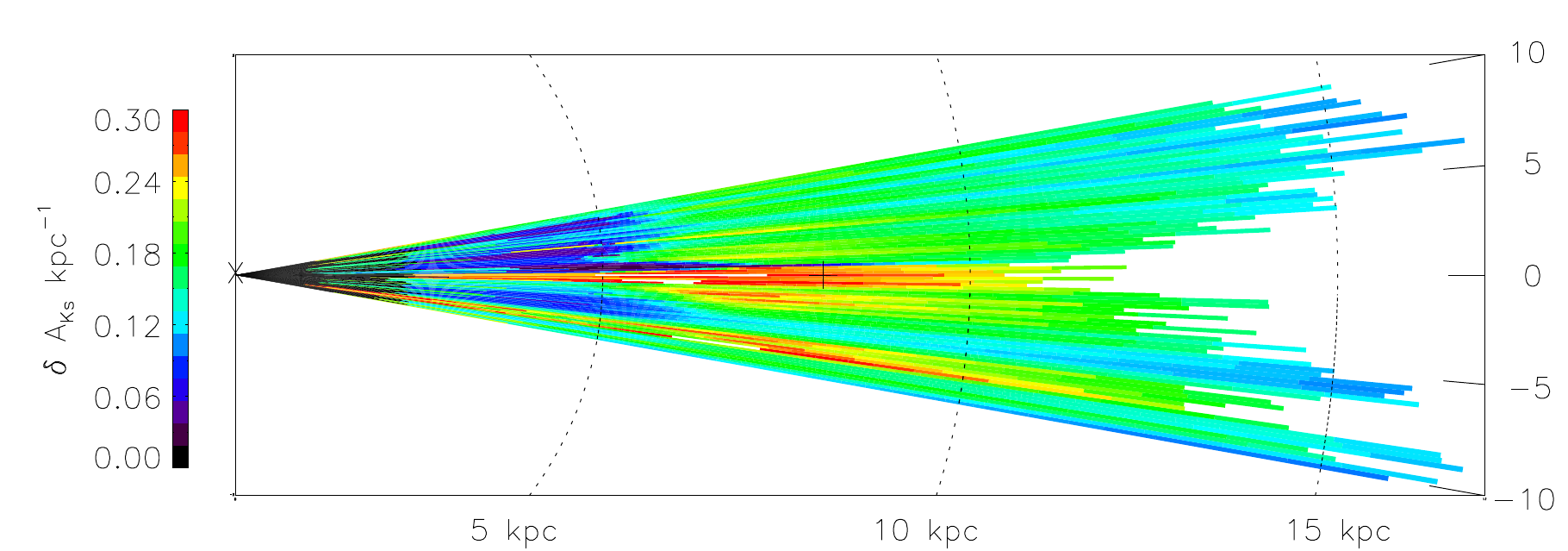}
   \includegraphics[width=18cm]{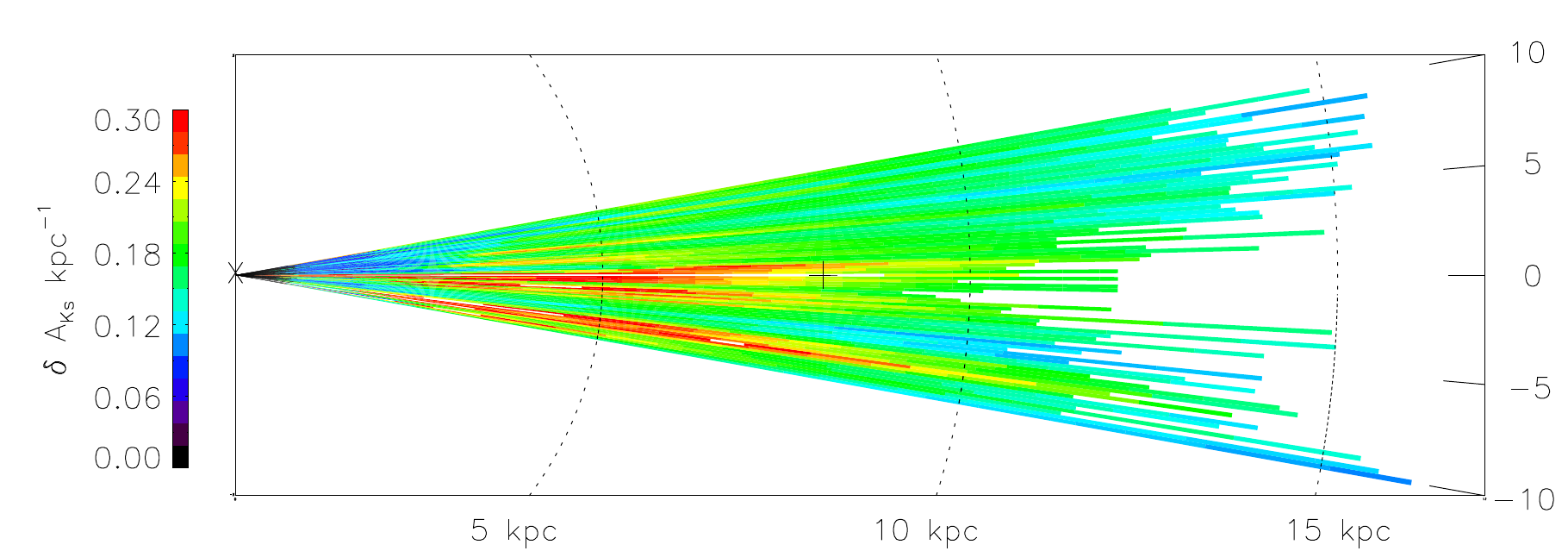}
\caption{Dust extinction in the Galactic Plane for $\rm |b| < 0.25$. The dotted lines correspond  to  distances of  5, 10 and 15\,kpc. The Sun position is marked as ``X'' and the GC as ``+''. Indicated on the y-axis is the galactic longitude. The upper panel shows our results while the lower one those from \citet{marshall2006}.}
\label{3dplot}
\end{figure*}

 \begin{figure}[!htbp]

   \includegraphics[width=9.0cm]{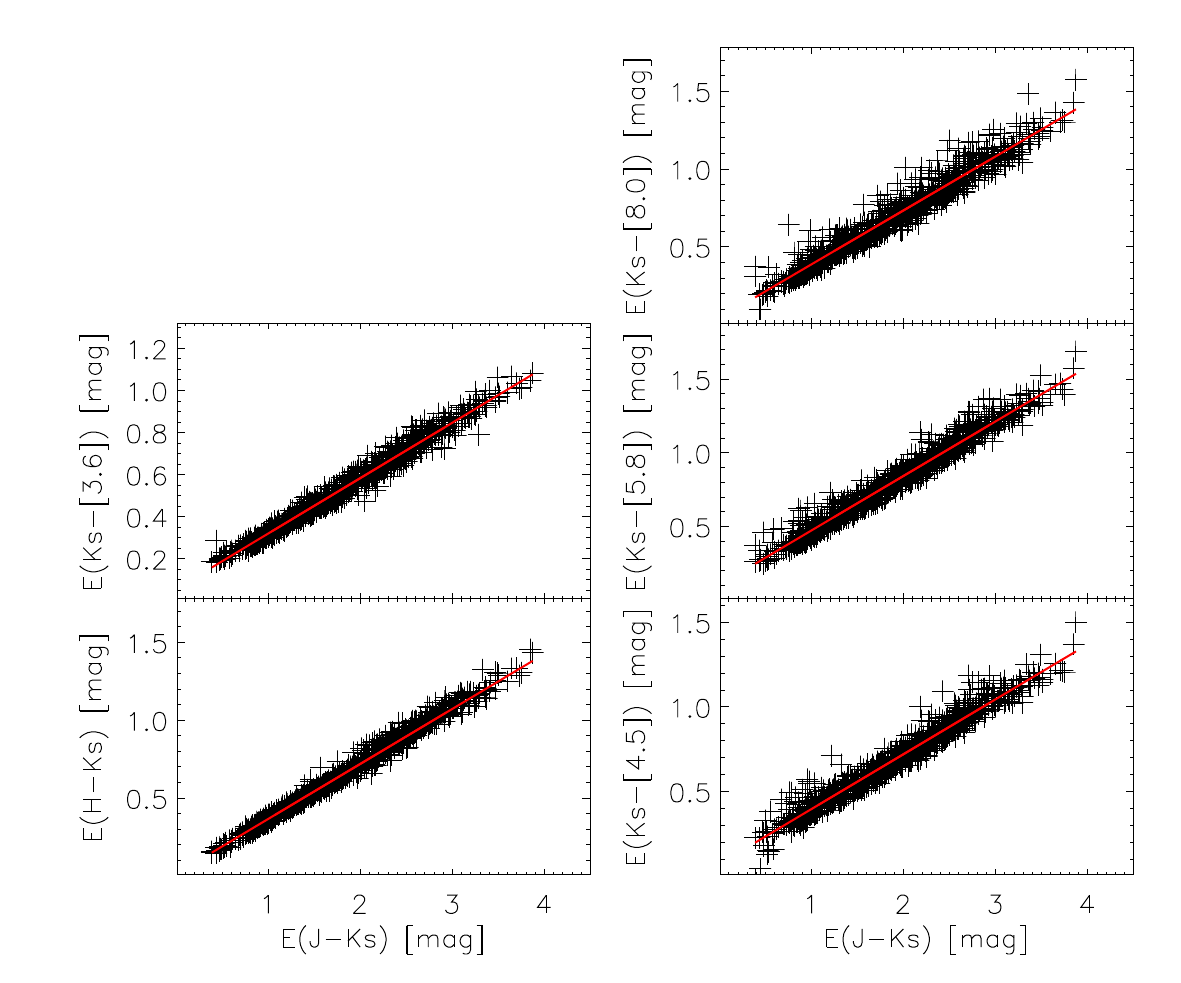}
\caption{Comparison of derived color excess values from J--Ks (VVV) to
 H--Ks (VVV), Ks--[3.6] (GLIMPSE-II), Ks--[4.5] (GLIMPSE-II), Ks--[5.8] (GLIMPSE-II) and Ks--[8.0] (GLIMPSE-II).}
\label{fig24}
\end{figure}

As already mentioned in Sect.~\ref{completeness} the VVV data used here are only for Ks fainter than 12\,mag. For this reason, the first bin of the VVV data is 4\,kpc. The VVV data on the other hand goes much deeper (d $\sim$ 10--12\,kpc). But 2MASS starts to be incomplete at 12\,mag depending on the line of sight. Therefore, there is in general a very small overlap between 2MASS and VVV which is also seen in the CMDs (see Figs.~\ref{fig7} and \ref{fig8}) where we miss in the VVV  data the bright foreground branch of K giants clearly visible in GLIMPSE-II. For the four  [IRAC] bands we see very similar extinction vs. distance relations.

\begin{figure}[!htbp]
   \includegraphics[width=9.0cm]{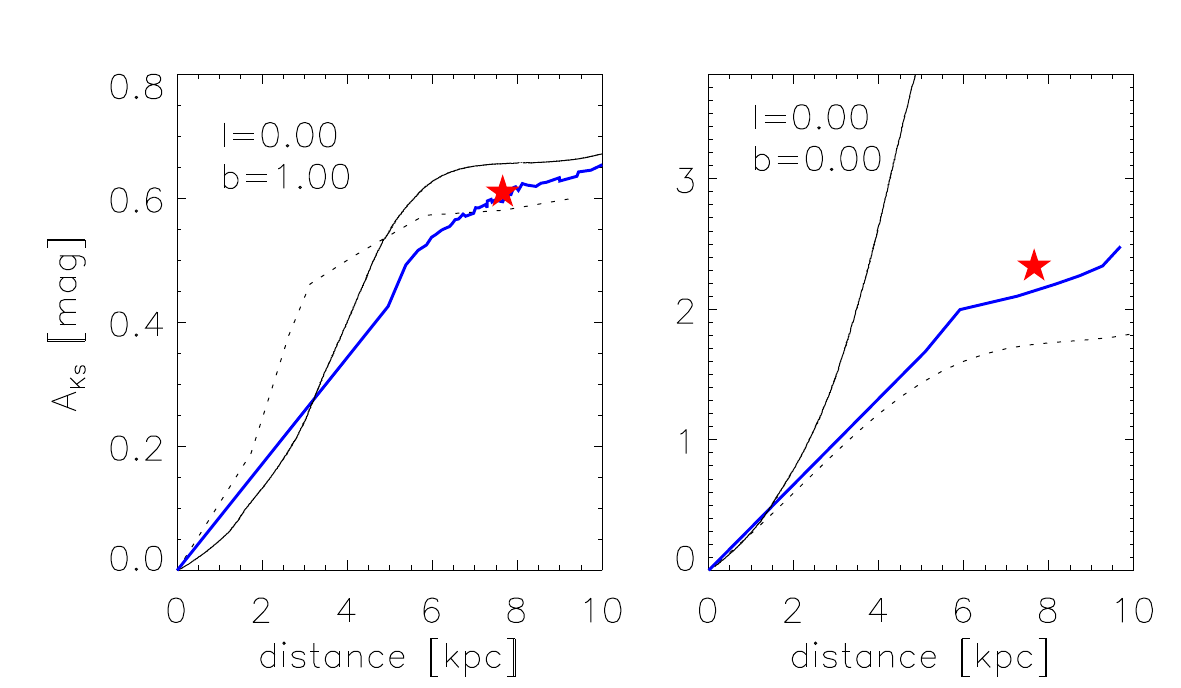}
\caption{3d extinction comparison with \citet{marshall2006} and \citet{drimmel2003}, where the black solid line is for \citet{drimmel2003} , black dashed line for \citet{marshall2006} and the blue line is the result of our determination. The star symbols denotes the calculated distance of the corresponding red-clump position of \citet{gonzalez2012}. }
\label{fig19}
\end{figure}

\subsection{The three dimensional extinction map}

In order to visualize the distribution of extinction in 3D we divided the extinction between subsequent bins by the distance between them similar as done by \citet{marshall2006}. The distance intervals are interpolated in  bins of 1\,kpc. All the 3D-maps using the six different color excesses i.e. $\rm E(J-Ks)$,  $\rm E(H-Ks)$, $\rm E(K-[3.6])$, $\rm E(K-[4.5])$, $\rm E(K-[5.8])$, $\rm E(K-[8.0])$  are presented in the Appendix~\ref{appa}. The units are in $\rm kpc^{-1}$. In addition we also show the error maps for each pair of filters. These errors are the combinations of the standard deviations of the observed colors and the synthetic ones. Note that on average the errors are lowest between 4--8\,kpc. We see a very similar picture of the dust distribution very concentrated to the galactic plane from all maps. As expected, extinction becomes higher as the distance from the Sun increases.

 We converted the seven color excess to $\rm A_{Ks}$ using the extinction law derived from Sect.~\ref{coeff} separately. The extinction map of the weighted mean values is shown in Fig.~\ref{fig9}. The maps are presented as extinction in Ks band and are also in units of $\rm kpc^{-1}$. In order to visualize better the dust extinction along the line of sight,  Fig.~\ref{3dplot} shows a view of our results (upper panel) together with that from \citet{marshall2006} (lower panel) from the North Galactic pole towards the galactic plane at $\rm |b| < 0.25$ similar to Fig. 9 of \citet{marshall2006}. Our results show globally  similar structures compared to \citet{marshall2006}. Except for $\rm l = 0^{0}$ and $\rm l= -7^{0}$ we notice beyond 6\,kpc  a very smooth and featureless dust extinction. Clearly seen  in this comparison  is also the much lower extinction until 6\,kpc compared to  \citet{marshall2006} which will be discussed in Sect.~\ref{comp3d}. The highest concentration of dust is around the Central Molecular zone, a giant molecular cloud complex with an asymmetric  distribution of molecular gas (see e.g. \citealt{morris1996}).  The dust extinction map shows very similar non-axisymmetric structure as observed in OH and CO. However, the spatial resolution of our  map is not high enough to trace features like the 100\,pc elliptical and twisted ring of cold dust discovered by \citet{molinari2011}.

The highest extincted regions are best traced by $\rm E(H-Ks)$ using the VVV data as well as $\rm E(Ks-[3.6])$ going up to $\rm A_{Ks} = 3.5\,mag$. 
Due to the high sensitivity of VVV, we are able to trace in 3D high extincted regions (such as star forming regions, molecular clouds,etc.)  until distances of 10\,kpc.

Concerning the $\rm E(Ks-[8.0])$ map we want to stress that at 8\,$\mu$m PAHs become an important component of emission which weakens the stellar flux (see e.g. \citealt{cotera2006}). We  get systematically higher extinction due to the PAHs emission.

Fig.~\ref{fig24} shows the comparison in the two-dimensional space between the color excess using   $\rm E(J-Ks)$  and the $\rm E(H-Ks)$, $\rm E(Ks-[3.6])$, $\rm E(Ks-[4.5])$, $\rm E(Ks-[5.8])$ and  $\rm E(Ks-[8.0])$ to see if we introduce any systematic offsets between the different filter pairs. The points with bootstrapping errors larger than 1$\sigma$ have been excluded.  In general we see a very good relation with a difference not larger than 0.1\,mag. The dispersion is higher for the GLIMPSE-II data  which is due to the larger photometric errors.

\begin{figure*}[htbp]
   \includegraphics[width=19cm]{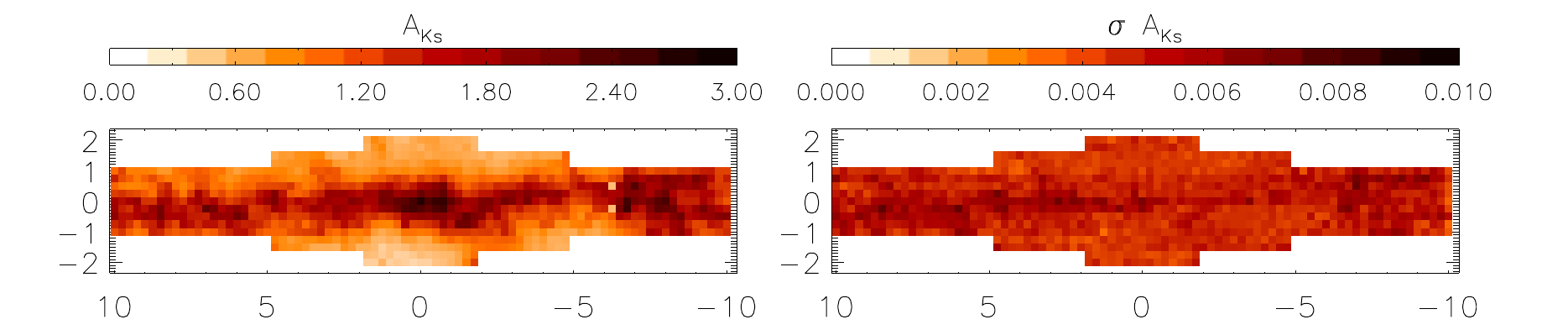}
\caption{2d extinction and error map for $\rm A_{Ks}$ integrated to a distance of 8\,kpc. The map shows the weighted average extinction converted from the seven individual colors excesses using the reddening law derived from Sect.~\ref{coeff}. X-axis denote galactic longitude while y-axis galactic latitude.}
\label{fig21}
\end{figure*}

\subsection{Comparison with other 3D-maps} \label{comp3d}

We compare  our results with that from \citet{marshall2006} and \citet{drimmel2003} for the extinction in the Ks band. Fig.~\ref{fig19}  shows the extinction v.s. distance diagram for two different fields, one located at  (l,b)=(0.00,1.00) and the other at extremely high extinction located at the Galactic Center (l=0.00, b=0.00) compared to the dust model of \citet{drimmel2003} and to  \citet{marshall2006}.

We can see that they show the same trend as mentioned in Sect.~5.1. The black solid line is from \citet{drimmel2003} and the dotted line  from \citet{marshall2006}. In order to compare  with Marshall et al. (2006), we converted their $\rm A_{Ks}$ values assuming the extinction coefficients of \citet{nishiyama2009}. Figure ~\ref{fig19} shows that  for our reference field (l,b)=(0,+1) we get a similar $\rm A_{Ks}$ vs. distance relation for d $>$ 6\,kpc while Marshall et al. predicts a rather steep slope  for $\rm d < 6\,kpc$. For the high extinction field at (l,b)=(0,0) there is a quite significant  difference. We obtain a steeper slope and we get systematically higher extinction values for d $>$ 4\,kpc. We added also for comparison a data point of the 2D map of \citet{gonzalez2012} on the diagram marked as star  symbol. The distance was calculated by taking the difference of the  dereddened mean magnitude of the red clump and the absolute magnitude of the red clump which we assume to be $\rm M_{Ks}=-1.55\,mag$. As one can see the estimated distances from \citet{gonzalez2012} agree remarkably well with our distances. \citet{schoedel2010} used adaptive optics observations of the central parsec of the Galactic center in the $\rm Ks$ band to map the extinction. They obtained a mean extinction value around Sagittarius $\rm A^{*}$ of $\rm A_{Ks}=2.46^{m}$ with a distance to the GC of $\rm R_{0} = 8.02\,kpc$, while \citet{fritz2011} derived  $\rm A_{Ks}=2.42^{m}$. Our derived value of  $\rm A_{Ks}=2.34^{m}$  is within the errors of the cited papers.

For low extinction fields, our results are in agreement with \citet{marshall2006}. We get only for larger distances slightly higher $\rm A_{Ks}$ values than Marshall et al. (2006). However, for the high extincted fields such as the GC (right panel of Fig.~\ref{fig19}), we notice rather significant differences: While the extinction seems to flatten at a distance of about 6\,kpc for \citet{marshall2006}, we notice a steady increase in extinction reaching values of $\rm A_{Ks}= 2.5^{m}$ compared to only $\rm A_{Ks}= 1.5^{m}$ for Marshall et al. (2006). This low limit is due to the large incompleteness of 2MASS in this high extincted region compared to VVV. We emphasize that compared to \citet{marshall2006} our distance steps shown here are much smaller and we trace thus better the derived distance-color relations.

Thanks to its deeper photometry and higher spatial resolution, the  VVV survey catalogue has stars at larger distances than 2MASS. For a red clump star with $\rm M_{Ks} = -1.65^{m}$ as an example, a brightness at $\rm Ks = 12^{m}$, the reliable measurement by 2MASS towards crowded regions, means a distance of about 3.5\,kpc with $\rm A_{Ks}$=1\,mag, and 5.5\,kpc if no extinction is applied. Since the Marshall results are derived based on the 2MASS data, there are difficulties in probing far and highly extincted regions.  While at Ks=15 mag in the case of the VVV survey, the red clump star can be detected at a distance of 13.4 kpc if $\rm A_{Ks}$=1\,mag and 5.4\, kpc if $\rm A_{Ks}$=3\,mag. Considering the RGB and AGB stars that are mostly  brighter than red clump stars, the VVV survey detected many stars at a distance further than 5\,kpc. This depth of VVV makes the results at larger distance more solid. This also explains the systematically higher extinction we obtained for the GC field at l=0.0 and b=0.0 (see Fig.~\ref{fig19}) compared to  \citet{marshall2006}.

Our results are clearly very different from those of \citet{drimmel2003}, in particular for the Galactic Center region. The Drimmel model predicts much higher extinction than ours as well as Marshall's. The dust model of  \citet{drimmel2003} depends very much on the dust temperature. However, in the Galactic Center region the gas pressure and temperature  is higher than in the galactic disk (see  \citealt{Serabyn1996}) which explains the large difference in the extinction values compared to  \citet{drimmel2003}. Our map is not sensitive to dust temperature like theirs but to the modeled K/M giant stellar population.
As our map is restricted to a spatial resolution of 15\arcmin $\times$ 15\arcmin, small-scale variations such as seen e.g. by \citet{gosling2010}  cannot be resolved.

\subsection{Two dimensional extinction maps}

We compare our extinction maps with the 2D-maps of  \citet{schultheis1999} and \citet{gonzalez2012}. While \citet{schultheis1999}  used mainly the RGB/AGB star population,  \citet{gonzalez2012} restricted themselves  to the red clump stars to trace the extinction.  For that purpose we integrate our extinction map up to a distance to 8\,kpc. The resulting 2D-maps of the combined $\rm A_{Ks}$ are shown in Fig.~\ref{fig21}. The individual maps of the different color excesses are presented in the Appendix~\ref{appb}. They show clearly very similar dust features as described in \citet{schultheis1999} and \citet{gonzalez2012}. We see clearly the small scale variation of the interstellar dust clouds concentrated towards the galactic plane which is not symmetric. Besides the clear dust features of the Central Molecular Zone,we also see a clear correlation between the high extinction and the number of detected infrared dark clouds based on molecular line observations (\citealt{Jackson2008}), especially at the region around (l,b)=(353.0, 0.0) with a large concentration of YSOs.

\begin{figure}

   \includegraphics[width=8cm]{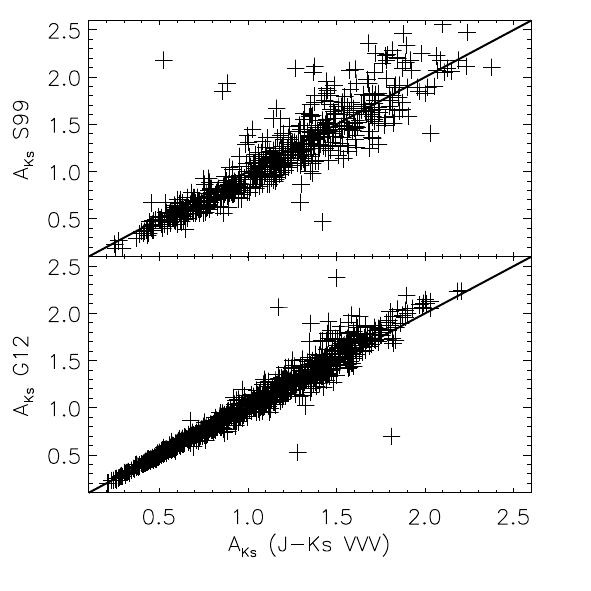}
\caption{Comparison of $A_{Ks}$ with \citet{schultheis1999} (top panel) and \citet{gonzalez2012} (lower panel).}
\label{fig25}
\end{figure}

For a more quantitative comparison we smoothed the \citet{gonzalez2012} maps to the same spatial resolution as ours (15\arcmin $\times$ 15\arcmin) and corrected for differences in the extinction law. Figure~\ref{fig25} shows the comparison. We see an excellent agreement between these maps with no systematic offset. For $\rm A_{Ks} > 1.5 \,mag$ the dispersion between our map and that of \citet{schultheis1999} gets larger which is caused mainly by the limited sensitivity of the DENIS survey in the J and Ks bands. In contrast, the dispersion compared to the VVV extinction is rather small, even for higher $\rm A_{K}$. We are thus confident that our extinction determination is highly reliable.

\begin{figure}

   \includegraphics[width=8cm]{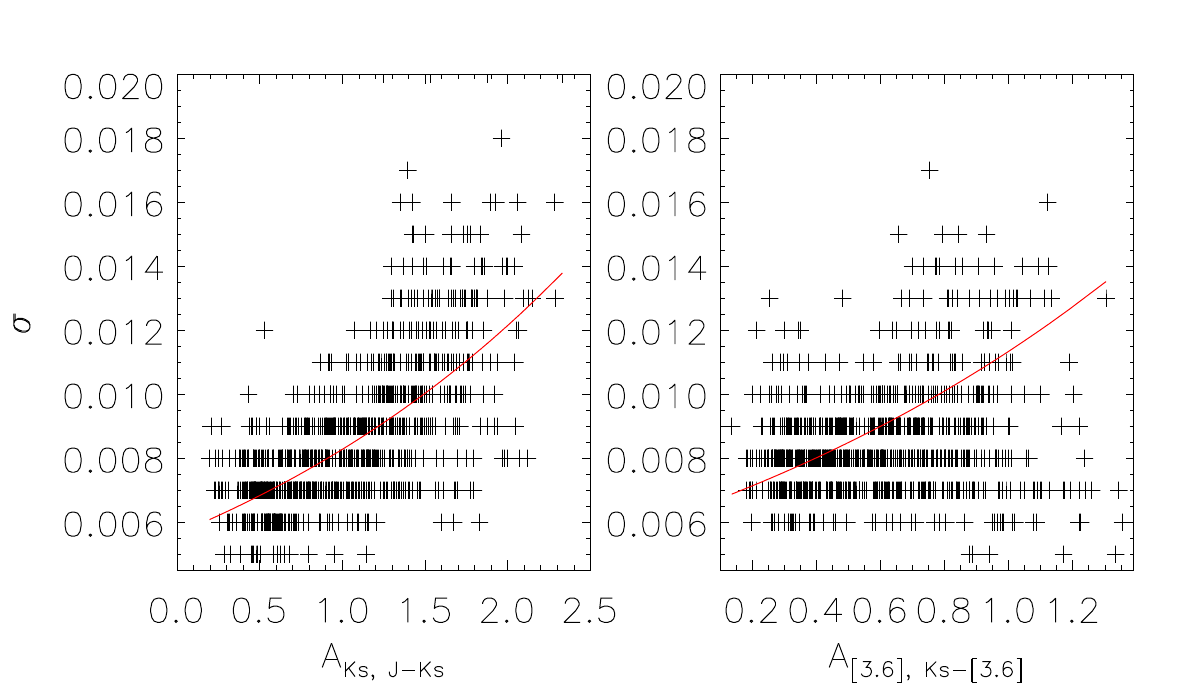}

\caption{The errors v.s. the extinction diagram for VVV (left panel) and 2MASS+GLIMPSE (right panel). The red solid line is an exponential fit.}
\label{fig22}
\end{figure}

Figure.~\ref{fig22}   shows that the intrinsic dispersion grows with increasing extinction. \citet{Lada1994} demonstrated that the form of the observed
$\rm \sigma_{disp}$ versus $A_{V}$ relation can be used to place constraints on the nature of the spatial distribution of extinction. However, photometric uncertainties as well as uncertainties in the extinction law dominate this relation. Note that the dispersion in VVV is much smaller than for 2MASS due to its much smaller photometric errors.
A higher spatial resolution would be needed to study this feature more in detail. The 2D maps of \citet{gonzalez2012} have 2\arcmin $\times$ 2\arcmin resolution resolution but their dispersion is dominated by the distribution of dust along the distance in that line of sight.

\begin{figure}
\begin{center}

   \includegraphics[width=6.5cm]{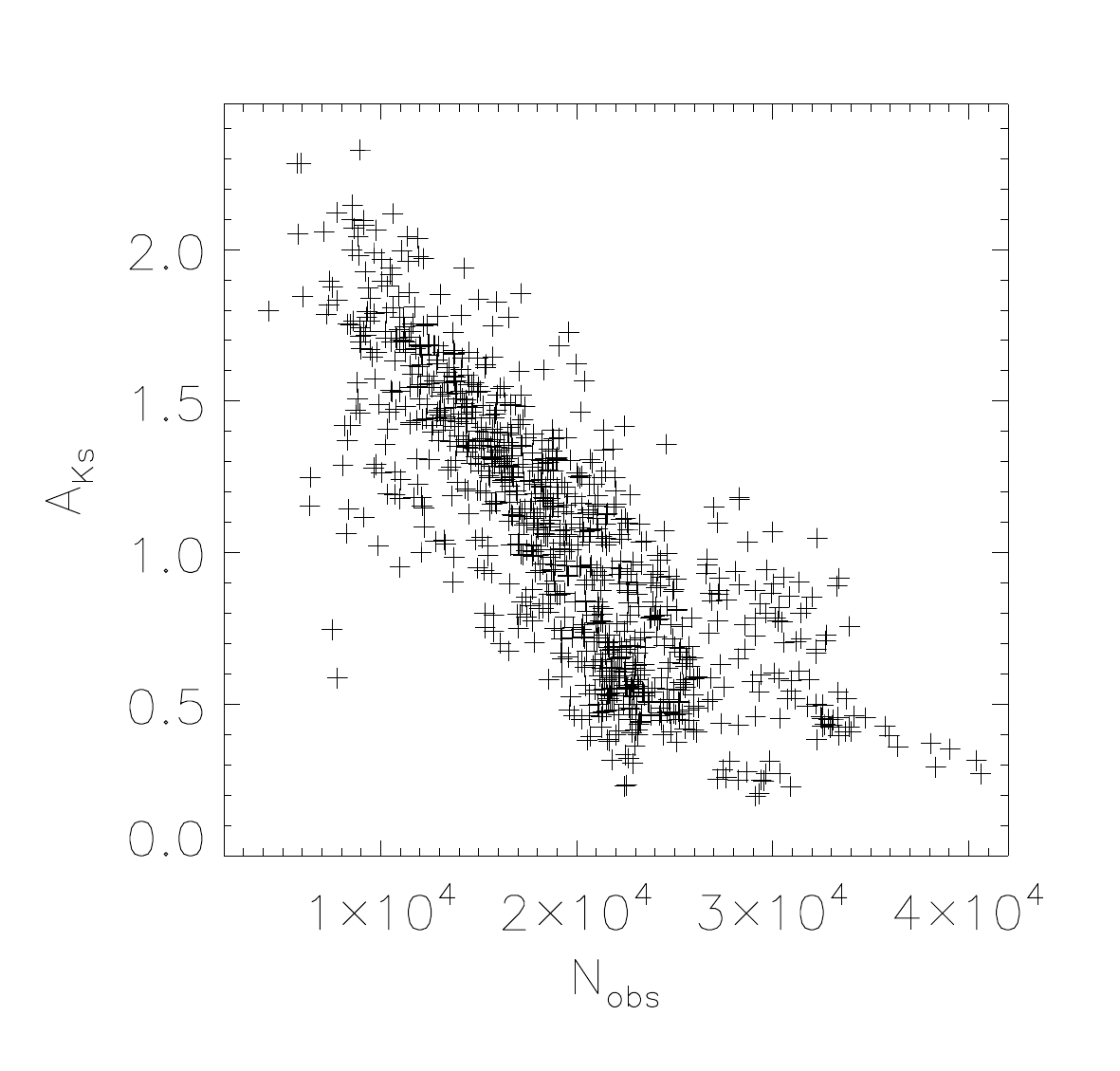}

\caption{Total number of stars for each VVV field  as a function of $\rm A_{K_{S}}$.}
\label{nobs}
\end{center}
\end{figure}

 Figure.~\ref{nobs}  shows  the total number of observed  stars for each VVV-subfield   and the corresponding $\rm A_{K_{S}}$.
We see clearly that the number  of stars gives an indication of the extinction, i.e., lower numbers means higher extinction.

\section{The extinction coefficients} \label{coeff}
As we determine the extinction maps using $E(\lambda - Ks)$ we can calculate the extinction coefficients. For the extinction coefficients $A_\lambda / A_{Ks}$, we used here the values from \citet{nishiyama2009} as the initial values.  If $\lambda$ is the wavelength we are going to derive and  $\alpha$ is the assumed wavelength,  we can write:

\begin{figure}[!htbp]
   \includegraphics[width=8cm]{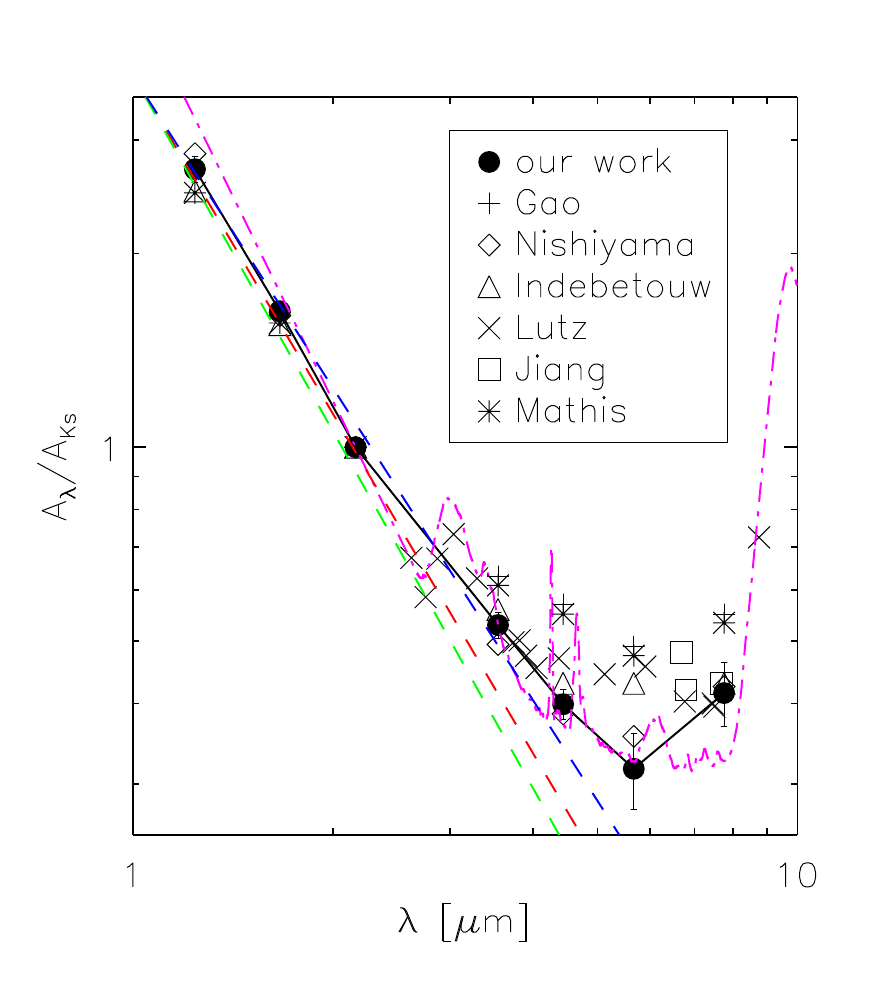}
\caption{Our derived interstellar extinction curve together with the comparison of different literature work. Note that we use the mean extinction values over all fields. The red dashed line indicates  a simple power law $A_\lambda \propto \lambda ^{-1.75}$ from \citet{draine1989} while the blue the power law of \citet{cardelli1989} with  $A_\lambda \propto \lambda ^{-1.61}$ and the green the power law of \citet{fritz2009} with  $A_\lambda \propto \lambda ^{-1.84}$. The dot-dashed line in magenta shows the extinction curve of \citet{fritz2011} in the  GC. }
\label{fig26}
\end{figure}

\begin{equation}
 A_\lambda / A_{Ks} = 1+(A_\alpha / A_{Ks} -1) \times  \frac{E(\lambda - Ks)}{E(\alpha -Ks)}
\end{equation}

$\rm A_\alpha / A_{Ks}$ is the initial value for the assumed wavelength from \citet{nishiyama2009}. $\rm E(\lambda - Ks)$ /($ E(\alpha - Ks)$) is the color excess derived from the iteration of $\chi ^2$ test (see Sect.4). By Eq.(3), we make the assumed wavelength as J, H, [3.6],[4.5],[5.8],[8.0] to calculate the extinction coefficients for each sub-fields. The values we adopted are from  \citet{nishiyama2009} because their line of sight of study is towards the Galactic center. However, considering the variation of the extinction law which are suggested by a couple of studies, using a unique value of $\rm A_{\alpha}/A_{Ks}$ could introduce some errors.

\begin{table*}[!htbp]
\caption{Extinction coefficients $A_\lambda / A_{Ks}$. }
\centering
\begin{tabular}{c c c c c c c c}
\hline\hline                        
$\alpha$	&	J	&	H & Ks &	[3.6]	& [4.5]	&	[5.8]	&	[8.0] \\
 $\lambda (\mu m)  $       &    1.24* & 1.664* & 2.164* & 3.545 & 4.442 & 5.675 & 7.760 \\
\hline
 &  &  & VVV & & & & \\
\hline
$\rm fixed\,J$  &     \bf{ 2.86 $\pm$0.01 }  &	1.68 $\pm$0.03  & 1.00 &     - &   -  &  -  &	- \\
$\rm fixed\,H$  &       2.63 $\pm$0.18  &  \bf{ 1.60 $\pm$0.01} &	  1.00   &  -  &  - &    -  &	- \\
\hline
 &  &  & GLIMPSE II & & & & \\
\hline
$\rm fixed\,[3.6]$   &       2.76 $\pm$0.10  &   1.66 $\pm$0.03  &   1.00 &  \bf{ 0.49 $\pm$0.01 } &   0.38 $\pm$0.02  &    0.28 $\pm$0.03  &	0.37 $\pm$0.04 \\
$\rm fixed\,[4.5]$   &       2.75 $\pm$0.15  &	  1.65 $\pm$0.05  &   1.00 &  0.50 $\pm$0.04  &  \bf{ 0.39 $\pm$0.01}  &    0.28 $\pm$0.07  &	0.37 $\pm$0.08 \\
$\rm fixed\,[5.8]$   &       2.57 $\pm$0.11  &	  1.58 $\pm$0.03  &  1.00 &   0.55 $\pm$0.01  &   0.45 $\pm$0.02  &   \bf{ 0.36 $\pm$0.01 } &	0.44 $\pm$0.03 \\
$\rm fixed\,[8.0]$   &       2.61 $\pm$0.11  &	  1.60 $\pm$0.03  &   1.00 &  0.54 $\pm$0.03  &   0.43 $\pm$0.03  &    0.33 $\pm$0.04  &	\bf{0.42 $\pm$0.01 }\\
\hline
mean &        2.70$\pm$0.13 &  1.63$\pm$ 0.03 &     1.00 &  0.53$\pm$0.02 &    0.40$\pm$0.02   &  0.32$\pm$0.04  &   0.42$\pm$0.05 \\
Indebetouw &  2.50$\pm$ 0.15 & 1.55$\pm$ 0.08 & 1.00 & 0.56$\pm$ 0.06 & 0.43 $\pm$ 0.08 & 0.43$\pm$ 0.10 & 0.43$\pm$ 0.10 \\
Nishiyama &      2.86  $\pm$0.08 &  1.60 $\pm$0.04 & 1.00 &  0.49$\pm$0.01 &  0.39$\pm$0.01 & 0.36$\pm$0.01   & 0.42$\pm$0.01 \\
Gao  &  - & - & 1.00 & 0.63$\pm$ 0.01 & 0.57$\pm$ 0.03 & 0.49$\pm$ 0.03& 0.55$\pm$ 0.03 \\
\hline
\end{tabular}
\tablefoot{The center wavelength are those of the 2MASS filters. In bold face  the assumed extinction $A_{\alpha} / A_{Ks}$ is indicated. }
\end{table*}

\subsection{The mean extinction coefficients}
Different choices of the assumed $A_{\alpha}/A_{Ks}$ would get different results for $A_{\lambda}/A_{Ks}$. We list our results in Table~3 varying the assumed extinction coefficient  $A_\alpha / A_{Ks}$. The mean extinction coefficients of all our fields are also listed in Table~3 and compared to \citet{gao2009},  \citet{indebetouw2005} and \citet{nishiyama2009}.
 The assumed values are noted as bold fonts in Tab.~3. Figure \ref{fig26} shows the derived extinction curve. The black circles indicate our mean value (see Tab. 3). 
 The red line indicates a simple power law $A_\lambda \propto \lambda ^{-1.75}$ from \citet{draine1989} while the blue line the power law of \citet{cardelli1989} with  $A_\lambda \propto \lambda ^{-1.61}$ and the green line the power law of \citet{fritz2009} with  $A_\lambda \propto \lambda ^{-1.84}$. It is clear that these simple power laws can not explain the flattening of the extinction curve between 3 -- 8\,$\mu$m. As shown by \citet{gao2009} (see their Fig. 7) a larger mean dust grain size (e.g. 0.3\,$\mu$m)  would give a steeper extinction law (i.e., smaller $A_{\lambda }/A_{{K_{s}}}$ ratios) than that of the diffuse ISM where the  mean dust size is 0.1\,$\mu$m. 
Our derived extinction curve is very similar to that of \citet{nishiyama2009}. Interestingly, we find  a smaller $\rm A_{[5.8]}/A_{Ks}$ value than   \citet{nishiyama2009}.  However,  our value is within the errors close to the value of \citet{nishiyama2009}. \citet{fritz2011} studied the near-IR and mid-IR extinction law towards $\rm Sgr\,A^{*}$ using hydrogen emission lines. They found a power-law slope of $\rm alpha  = -2.11 \pm 0.06$ shortward of 2.8\,$\mu$m. This is in agreement with our derived value of $\rm \alpha = -2.175$ and those found in the literature. Figure \ref{fig26} shows that our derived extinction curve agrees remarkably well to that
 of  \citet{fritz2011} (indicated by the dot-dashed line) and confirms our low extinction value at [5.8]. As pointed out by  \citet{fritz2011} classical grain models  (\citealt{Li2001}, \citealt{weingartner2001}).
 using mainly silicate and graphite grains with different size distributions fail to reproduce the observed extinction law and especially the change of slope in the extinction law. Water ice features could be responsible for the slope change in the mid-IR. However further modeling is necessary.

The comparison shows that our mean values in the IRAC bands are comparable to those  found in the literature.  However it depends on the assumed $\rm A_{\alpha}/A_{Ks}$. It can be seen from Table~3 that $\rm A_{\lambda}/A_{Ks}$ gets smaller value when $\rm A_{\alpha}/A_{Ks}$ is bigger, which is a natural result from equation (3) because ($\rm A_{\alpha}/A_{Ks}$-1) is negative for any wavelength longer than Ks. The values of $\rm A_{J}/A_{Ks}$ assumed or derived have a minimum of 2.56, a maximum of 2.89 and a mean of 2.70. In comparison, \citet{gao2009} used  $\rm A_{J}/A_{Ks}$=2.52. 

\begin{figure*}
   \includegraphics[width=18cm]{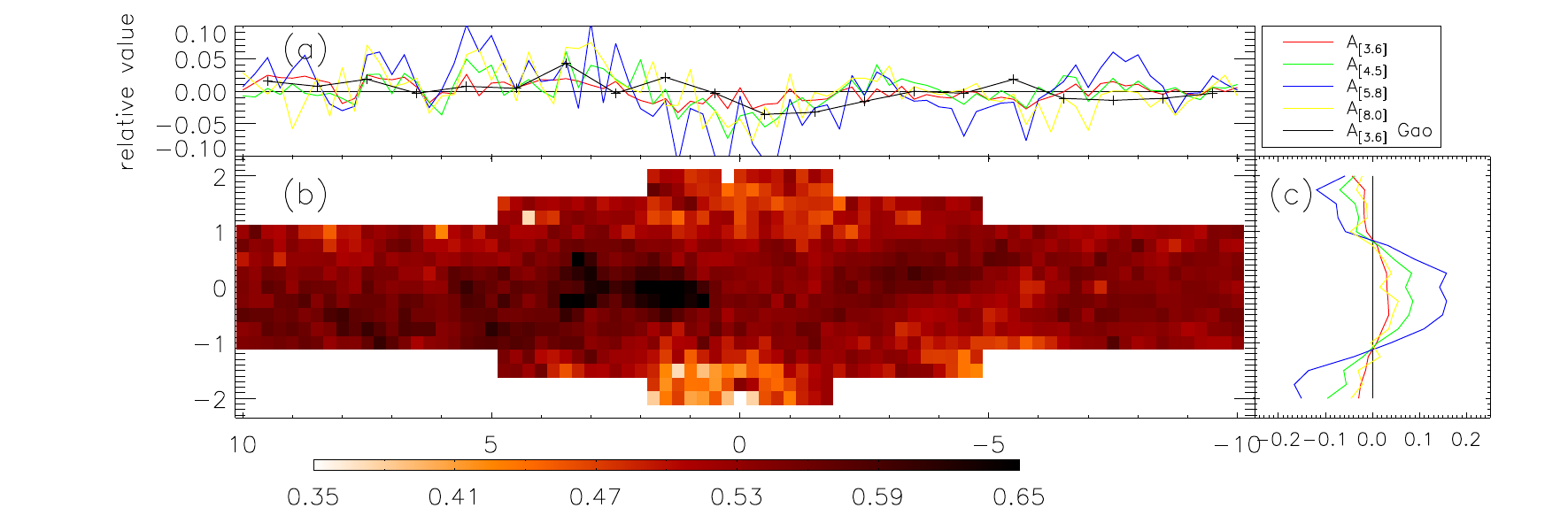}
\caption{Galactic longitude vs. galactic latitude map of the  $A_{[3.6]}/A_{Ks}$ extinction coefficient  assuming $A_J/A_{Ks}=2.86$ as long as longitudinal and latitudinal distribution of  $A_\lambda /A_{Ks}$ ($\lambda$ equals to  3.6, 4.5, 5.8 and 8.0 $\mu$m). (a): The distribution of $A_\lambda /A_{Ks}$ as function of the galactic longitude. Different colors denotes different wavelengths, the zero line stands for fixed $A_J/A_{Ks}$ and the black line indicates the data from \citet{gao2009}. (b): The map of the  $A_{[3.6]}/A_{Ks}$ extinction coefficient, x-axis denote galactic longitude while y-axis galactic latitude. (c): The same as (a) but for the galactic latitude.}
\label{fig27}
\end{figure*}

\subsection{Variation of the extinction coefficients}

The extinction coefficient may differ along  different lines of sight. \citet{gao2009} showed an interesting distribution of extinction ratios as a function of  galactic longitude with 131 GLIMPSE fields. In this work, we take the extinction coefficients $A_{\lambda}/A_{Ks}$ of all our sub-fields in our field. We take the median value of the $A_{\lambda}/A_{Ks}$ in each bin of longitude or latitude. To see the variation clearly, we present here the relative value  ($\frac{x-\bar{x}}{\bar{x}}$).  Figure \ref{fig27}b shows a 2D map of $A_{[3.6]}/A_{Ks}$  assuming $A_J/A_{Ks}=2.86$, and Fig. \ref{fig27}a,c show the variation of the extinction coefficients (i.e. $A_{[3.6]}/A_{Ks}$,$A_{[4.5]}/A_{Ks}$,$A_{[5.8]}/A_{Ks}$ and $A_{[8.0]}/A_{Ks}$) as a function of galactic longitude (Fig. \ref{fig27}a) and latitude (Fig. \ref{fig27}c). Overplotted in figure \ref{fig27}a are also the values from  \citet{gao2009}. Note that they use larger binsizes (1 degree) compared to ours. We notice in all [IRAC] bands a similar behavior in the variation of the extinction coefficient along the galactic longitude which  follows the extinction curve variation of \citet{gao2009}. While this variation is rather small, we see a surprisingly large peak visible towards the galactic plane indicating  a large variation with galactic latitude (see Fig. ~\ref{fig27}b). This variation seems to be higher for longer wavelengths  except for  8.0 $\mu$m where  the amplitude becomes small again.  This might be related to  the 9.7 $\mu$m silicate absorption feature \citep{gao2009}. Figure ~\ref{fig27} indicates that the extinction curve becomes flatter in the mid-IR approaching the galactic Center region.  A detailed comparison  with interstellar dust models is needed to explain this feature which is beyond the scope of this paper.

\section{Conclusion}

 Using an improved version of the Besan\c{c}on model, we present here 3D extinction maps in the J, H, Ks, [3.6], [4.5], [5.8] and [5.8] bands using GLIMPSE II  and VVV data. All the extinction maps are available  online at  CDS (Table 1 and Table 2), as well as  through the BEAMER webpage (http://mill.astro.puc.cl/BEAM/calculator.php). We derived new temperature -color relation for M giants which match better the observed color-magnitude diagrams.

We present the 3D maps from 1.25\,$\mu$m until 8\,$\mu$m. Due to the high sensitivity of the VVV data, we are able to trace large extinction until 10\,kpc. Our maps integrated along the line of sight up to 8\,kpc show an excellent agreement to the 2D-maps from \citet{schultheis2009} and \citet{gonzalez2012}. These maps show the same dust features and they show consistent $A_{Ks}$ values.

 Using the initial value from \citet{nishiyama2009}, we derived the mean extinction coefficient of our field in the seven bands, $A_J/A_{Ks}$ = 2.70$\pm$0.13, $A_H/A_{Ks}$ = 1.63$\pm$ 0.03, $A_{[3.6]}/A_{Ks}$ = 0.53$\pm$0.02, $A_{[4.5]}/A_{Ks}$ = 0.40$\pm$0.02, $A_{[5.8]}/A_{Ks}$ =   0.32$\pm$0.04, $A_{[8.0]}/A_{Ks}$ = 0.42$\pm$0.05. The variation of the coefficients indicates  that the wavelength dependence of interstellar extinction in the mid-IR varies from different lines of sight, which means that there is no ``universal'' IR extinction law. This is also in agreement with the study of Gao et al. (2009).

\begin{acknowledgements}
We want to thank the referee for his/her fruitful comments.    
We gratefully acknowledge use of data from the ESO Public Survey program ID 179.B-2002 taken with the VISTA telescope, data products from
the Cambridge Astronomical Survey Unit and funding from the FONDAP Center for Astrophysics 15010003, the BASAL CATA Center for Astrophysics
and Associated Technologies PFB-06, the MILENIO Milky Way Millennium Nucleus from the Ministry of
Economy’s ICM grant P07-021-F, and the FONDECYT the Proyecto FONDECYT Regular No. 1090213. We are grateful to the GLIMPSE-II project for providing access to the data. B.Q. Chen was supported from a scholarship of the China Scholarship Council (CSC).
\end{acknowledgements}

\bibliographystyle{aa}
\bibliography{3dextin_v3.1}

\Online

\begin{appendix}

\section{3D reddening maps} \label{appa}

\begin{figure*}[!htbp]
   \includegraphics[width=18cm]{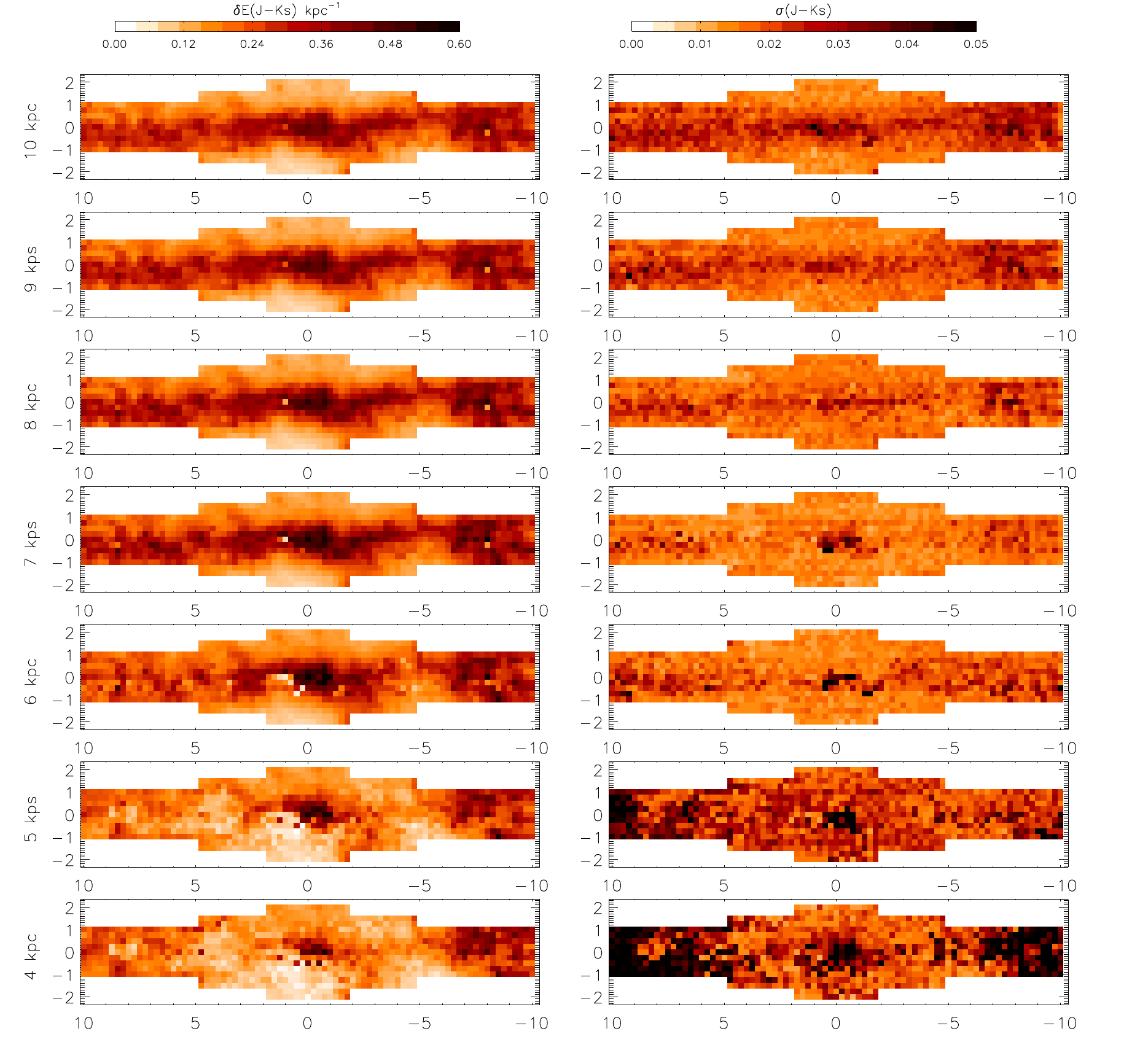}
\caption{3d reddening and error maps for J-Ks VVV. X-axis denote galactic longitude while y-axis galactic latitude. The units are in $\rm \delta E(J-Ks)\,kpc^{-1}$.}
\label{fig90}
\end{figure*}

\begin{figure*}
   \includegraphics[width=19cm]{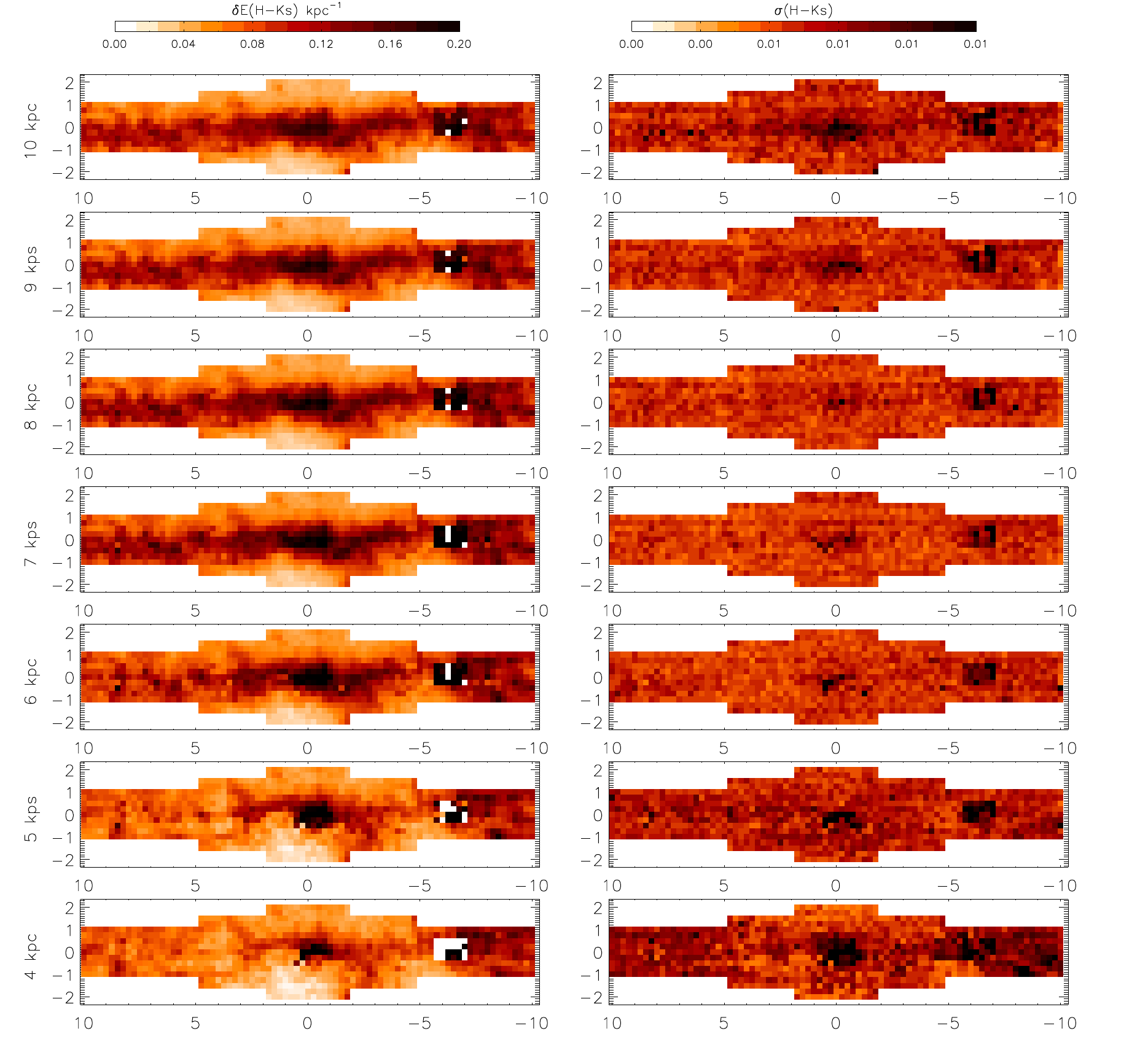}
\caption{3d reddening and error maps for E(H--Ks) VVV. }
\label{fig11}
\end{figure*}

\begin{figure*}
   \includegraphics[width=19cm]{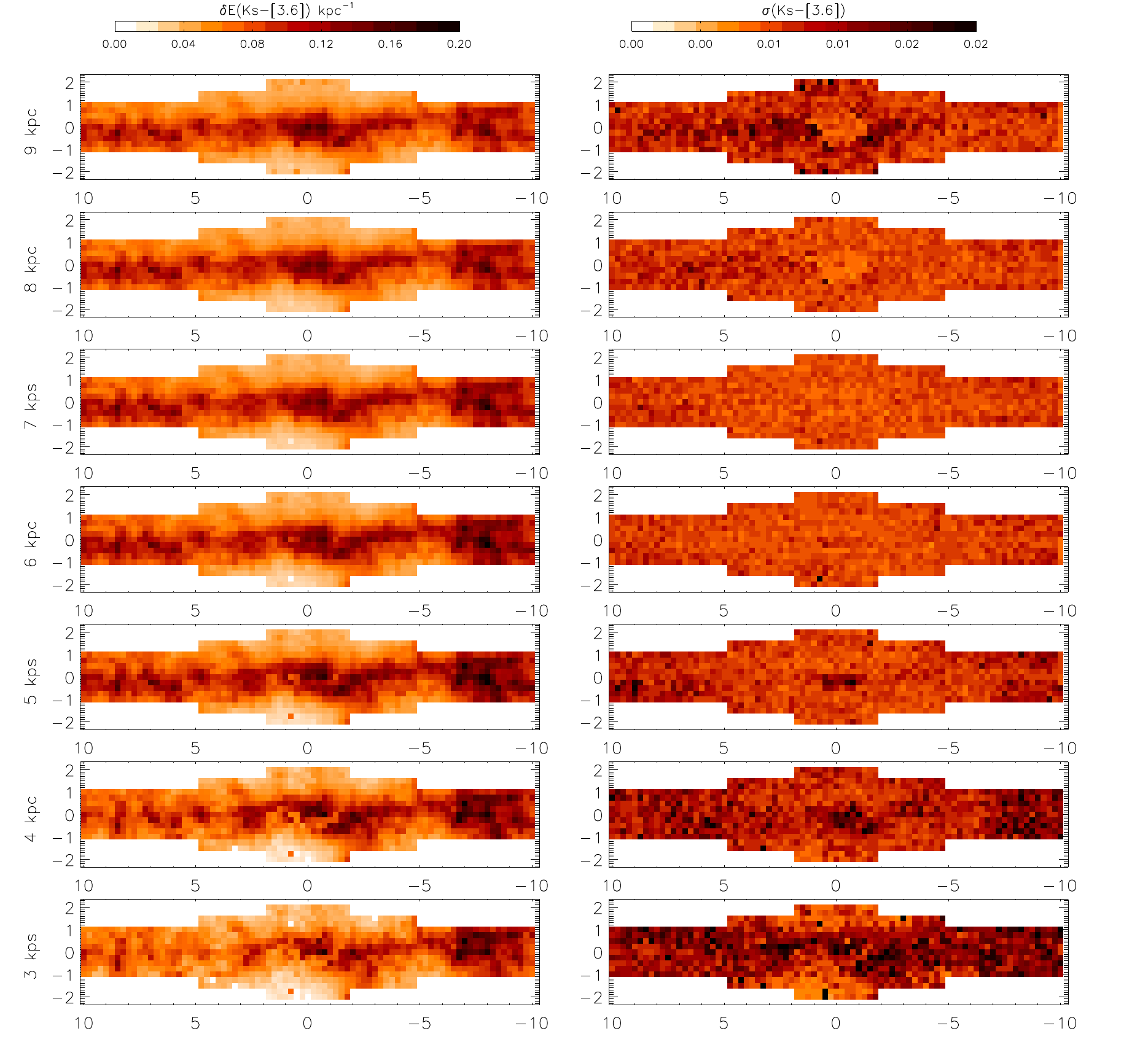}
\caption{3d reddening and error maps for E(Ks--[3.6]).}
\label{fig15}
\end{figure*}

\begin{figure*}
   \includegraphics[width=19cm]{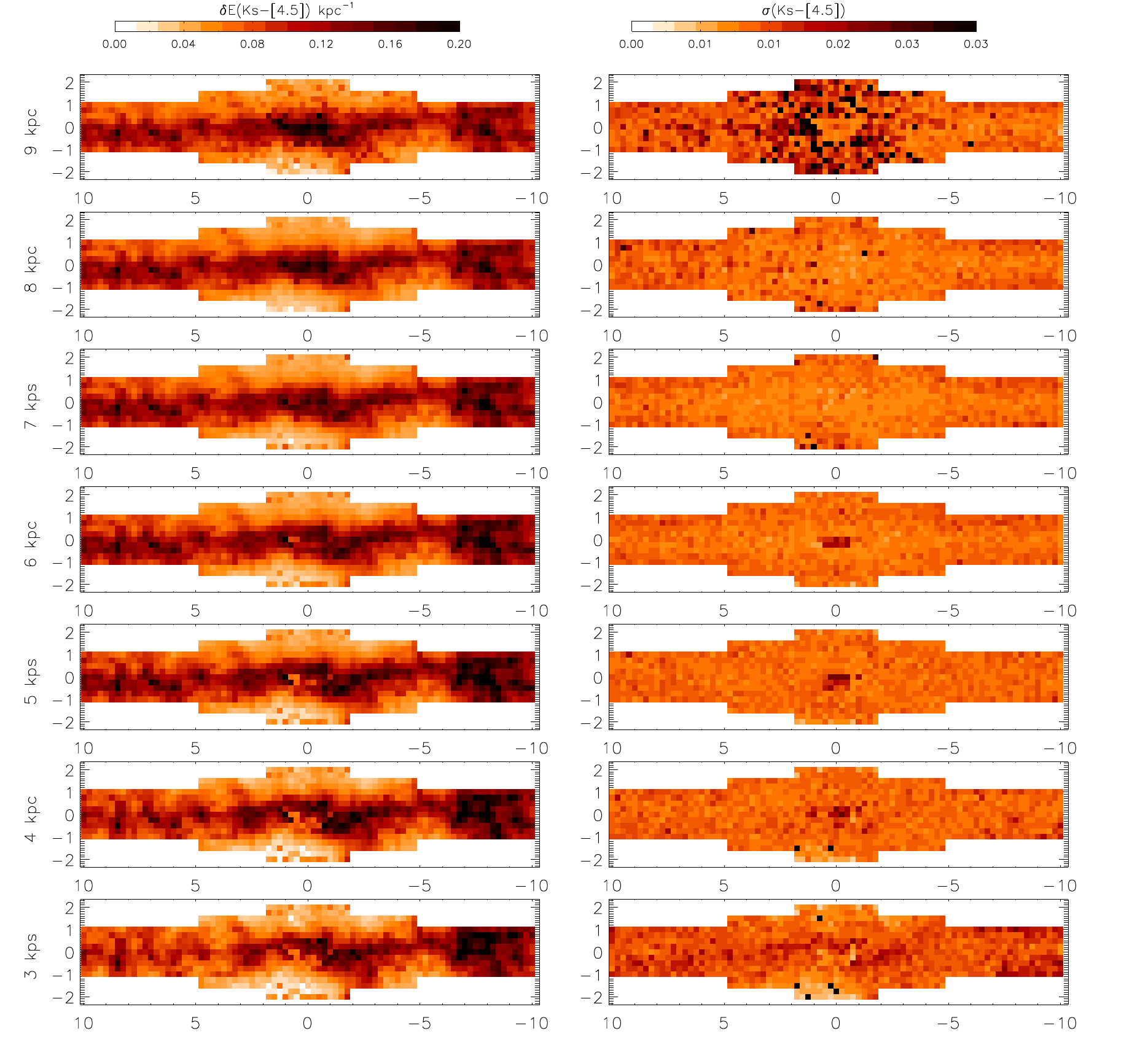}
\caption{3d reddening and error maps for E(Ks--[4.5]).}
\label{fig16}
\end{figure*}

\begin{figure*}
   \includegraphics[width=19cm]{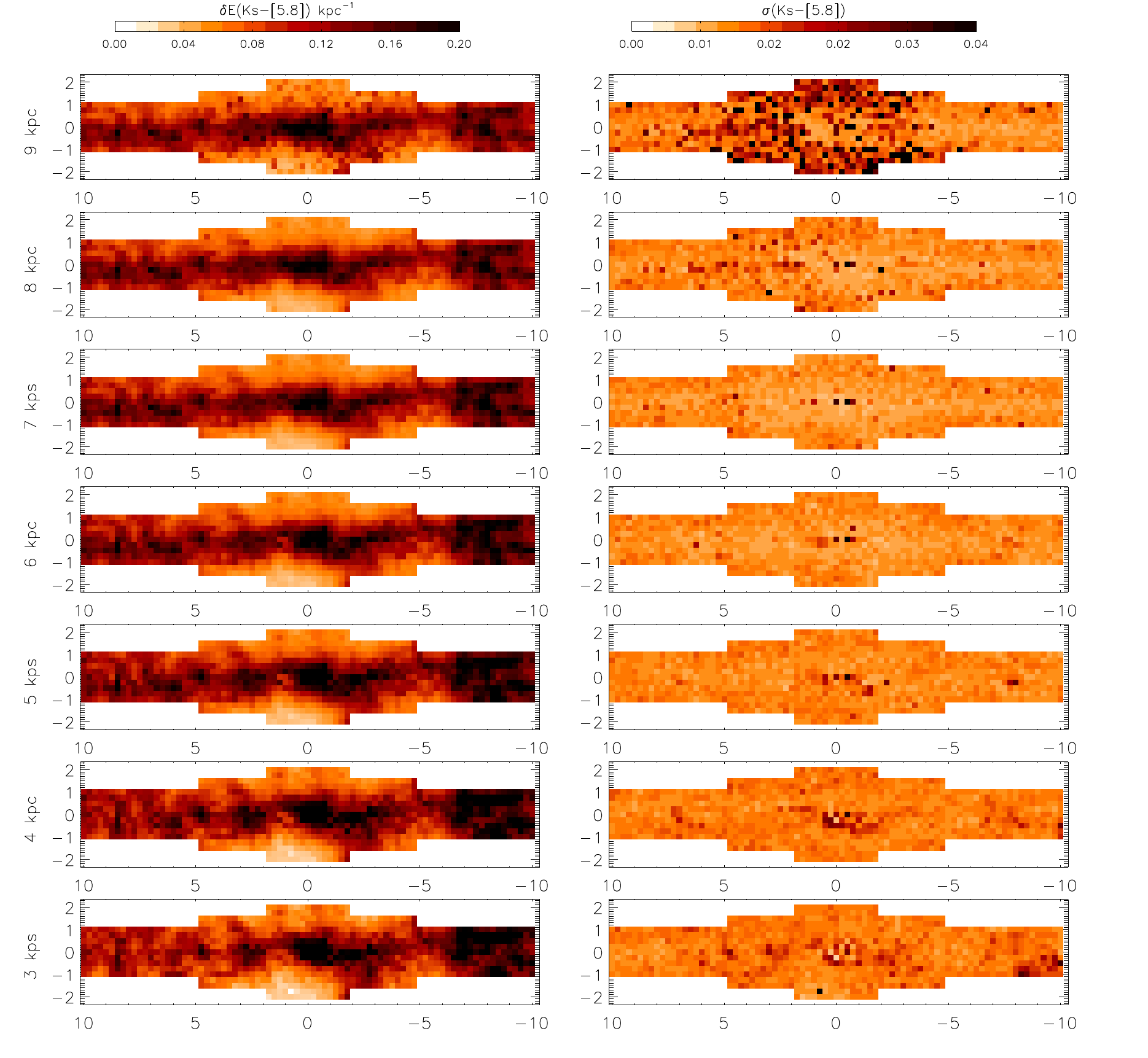}
\caption{3d reddening and error maps for E(Ks--[5.8]).}
\label{fig17}
\end{figure*}

\begin{figure*}
   \includegraphics[width=19cm]{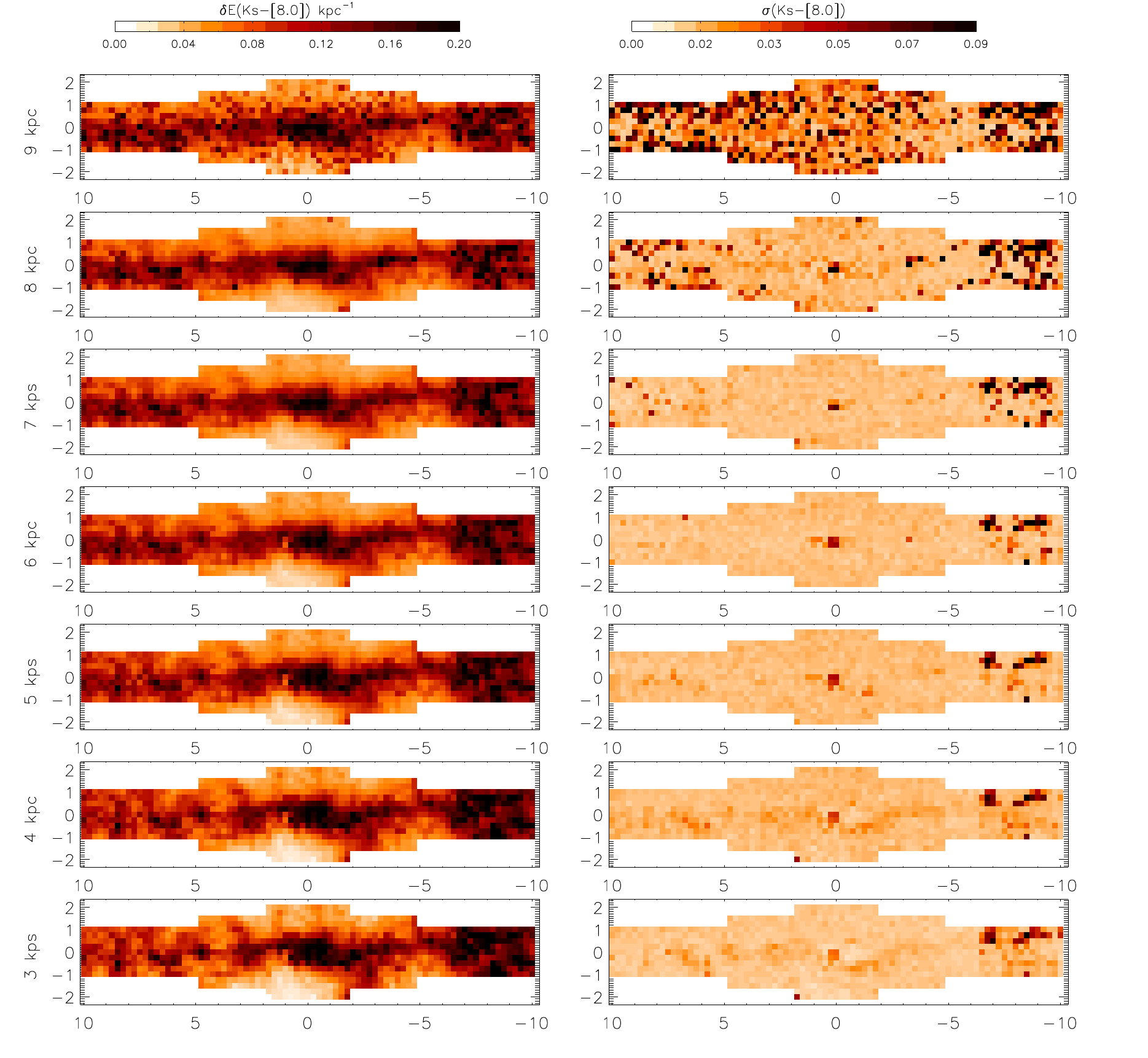}
\caption{3d reddening and error maps for E(Ks--[8.0]). }
\label{fig18}
\end{figure*}

\section{2D reddening maps} \label{appb}

\begin{figure*}[htbp]
   \includegraphics[width=19cm]{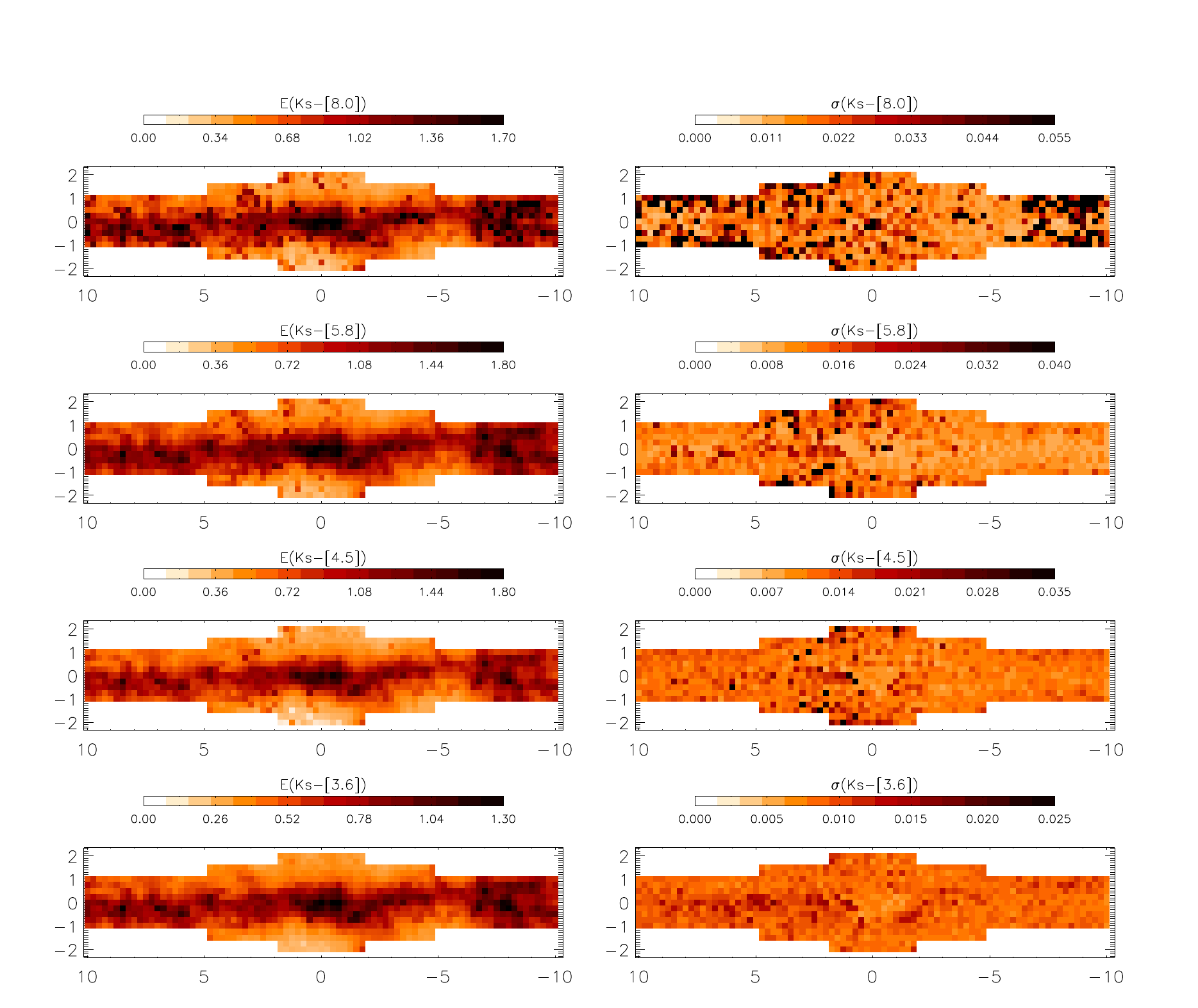}
\caption{2d reddening and error map for E(Ks--[3.6]), E(Ks--[4.5]), E(Ks--[5.8], E(K--[8.0]) integrated to 8\,kpc distance.}
\label{fig210}
\end{figure*}

\begin{figure*}[htbp]
   \includegraphics[width=17cm]{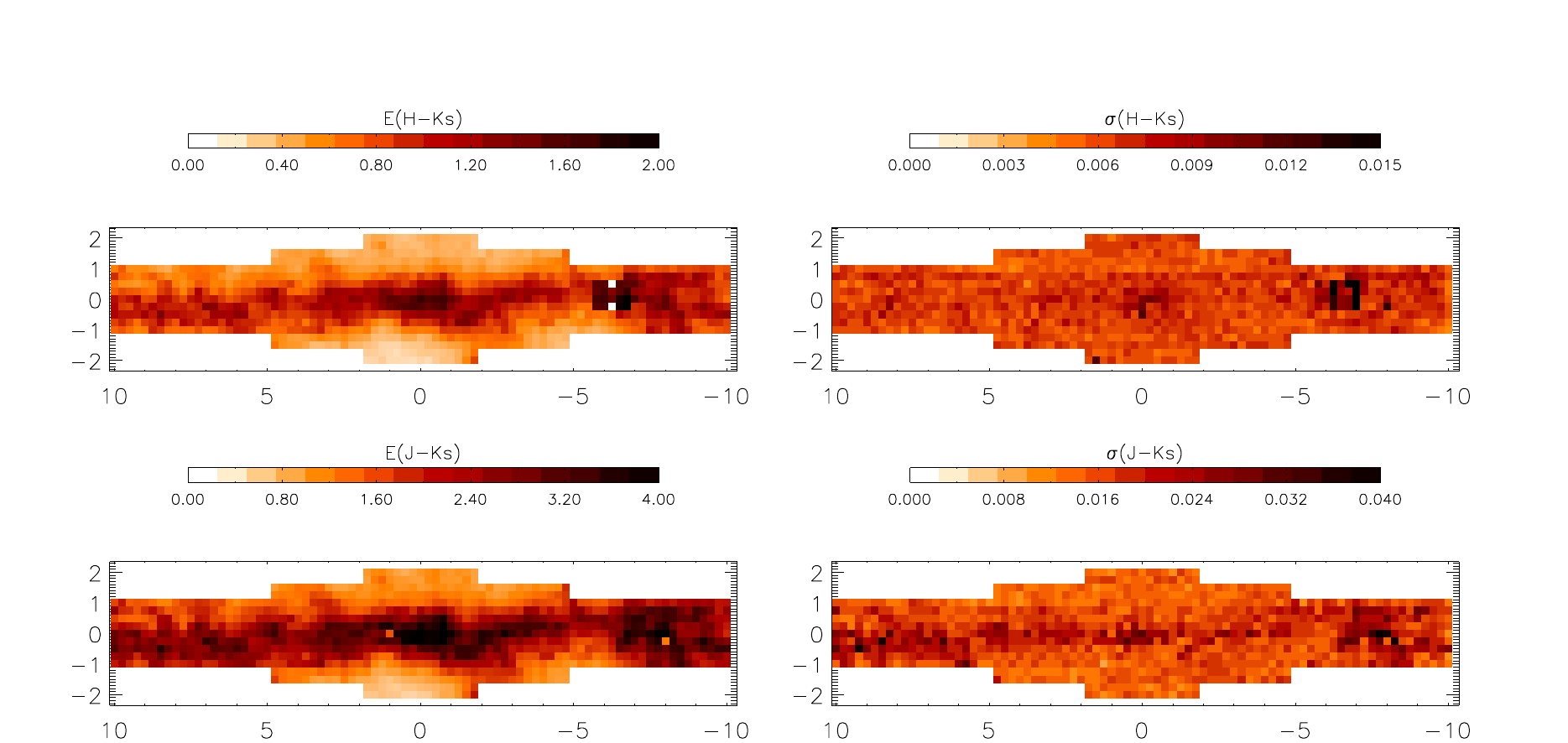}
\caption{2d reddening and error map for E(J--Ks), E(H--Ks) in the VVV bands integrated to 8\,kpc distance.}
\label{fig20}
\end{figure*}

\end{appendix}

\end{document}